\documentclass[a4paper,fleqn,usenatbib]{mnras}

\usepackage[T1]{fontenc}
\usepackage{ae,aecompl}

\usepackage{graphicx}	
\usepackage{amsmath}	
\usepackage{amssymb}	
\usepackage{pifont}
\usepackage{afterpage}
\usepackage{verbatim}
\usepackage{caption}
\usepackage{subfig, float}
\usepackage[dvipsnames]{xcolor}

\usepackage{newtxtext,newtxmath}

\title[QSO-host galaxies at high z]{Host galaxies of high-redshift quasars: SMBH growth and feedback}

\author[M. Valentini et al.]
{Milena Valentini$^{1,2,3,4}$\thanks{Alexander von Humboldt Research Fellow\newline E-mail: valentini@usm.lmu.de}, 
Simona Gallerani$^{3}$,
and Andrea Ferrara$^{3}$
\\ ~ \\
\footnotesize
$^{1}$ Universit{\"a}ts-Sternwarte,  Fakult{\"a}t f{\"u}r Physik, Ludwig-Maximilians-Universit{\"a}t M{\"u}nchen, Scheinerstr. 1, 81679 M{\"u}nchen, Germany\\
$^{2}$ Excellence Cluster ORIGINS, Boltzmannstr. 2, D-85748 Garching, Germany\\
$^{3}$ Scuola Normale Superiore, Piazza dei Cavalieri 7, I-56126 Pisa, Italy\\ 
$^{4}$ INAF - Osservatorio Astronomico di Trieste, via Tiepolo 11, I-34131 Trieste, Italy\\
}

\date{Accepted 2021 July 9.  Received 2021 June 21; in original form 2020 December 26.}

\pubyear{2018}

\begin{document}
\label{firstpage}
\pagerange{\pageref{firstpage}--\pageref{lastpage}}
\maketitle

\begin{abstract}
The properties of quasar-host galaxies might be determined by the growth and feedback of their supermassive (SMBH, $10^{8-10}$~M$_{\odot}$) black holes. We investigate such connection with a suite of cosmological simulations of massive (halo mass $\approx 10^{12}$~M$_{\odot}$) galaxies at $z\simeq 6$ which include a detailed sub-grid  multiphase gas and accretion model. BH seeds of initial mass $10^5$~M$_{\odot}$ grow mostly by gas accretion, and become SMBH by $z=6$ setting on the observed $M_{\rm BH} - M_{\star}$ relation without the need for a boost factor. Although quasar feedback crucially controls the SMBH growth, its impact on the properties of the host galaxy at $z=6$ is negligible. In our model, quasar activity can both quench (via gas heating) or enhance (by ISM over-pressurization) star formation. However, we find that the star formation history is insensitive to such modulation as it is largely dominated, at least at $z>6$, by cold gas accretion from the environment that cannot be hindered by the quasar energy deposition. Although quasar-driven outflows can achieve velocities $> 1000~\rm km~s^{-1}$, only $\approx 4$\% of the outflowing gas mass can actually escape from the host galaxy. These findings are only loosely constrained by available data, but can guide observational campaigns searching for signatures of quasar feedback in early galaxies. 
\end{abstract}




\section{Introduction} 
\label{sec:introduction}

Quasars are among the most luminous astrophysical sources: they shine at the centre of their host galaxies, 
where gas accretion fuels a supermassive black hole (SMBH), their engine. 
Quasar high luminosity allows their identification out to very-high redshift ($z \gtrsim 6$): 
they can be deemed as beacons in the early universe, thus being signposts of 
the early steges of galaxy evolution and black hole (BH) growth.  
More than $200$ quasars have been discovered over the last decades at $z \gtrsim 6$ 
by means of optical/near-infrared (IR) surveys 
\citep[e.g.][]{Fan2006, Willott2010, Mortlock2011, Venemans2013, Venemans2015, Banados2014, 
Jiang2016, Matsuoka2016, Matsuoka2019, Pons2019}, 
and have been studied through their ultraviolet (UV) and X-ray emission 
\citep[][]{Brandt2002, Farrah2004, Shemmer2006, Page2014, 
Gallerani2017, Koptelova2017, Nanni2017, Nanni2018, Salvestrini2019, Vito2019, Pons2020}. 
These observations have provided information about the physical properties 
of these powerful active galactic nuclei (AGN), characterized by bolometric luminosities 
$L_{\rm bol} \gtrsim 10^{46}$~erg/s 
\citep{Willott2010, DeRosa2014, Barnett2015, Venemans2015, Gallerani2017, Mazzucchelli2017, Matsuoka2019b}. 
In particular, it has been found that $z\sim 6$ quasars are powered by SMBHs 
with typical masses spanning the range $10^8 - 10^{10}$~M$_{\odot}$ 
\citep[e.g.][]{Ho2007, Wang2010, Venemans2016, Pensabene2020}. 
The advent of the Atacama Large Millimeter/submillimeter Array (ALMA) has later allowed to investigate 
the properties of the host galaxies of these distant quasars \citep[e.g.][]{Carilli2013, Wang2013, 
Venemans2016, Venemans2017, Venemans2017_845, Venemans2019, Gallerani2017PASA, Willott2017, 
Decarli2018, Feruglio2018, Carniani2019, Novak2019, Wang2019}.

The presence of SMBHs which grew as massive as $10^8 - 10^{10}$~M$_{\odot}$ 
in less than $\sim 1$~Gyr (i.e. the age of the universe at $z \simeq 6$) 
represents an important constraint for SMBH fromation channels, and 
poses a challenging question from a theoretical perspective 
\citep[see e.g.][for reviews, and references therein]{Volonteri2010, Volonteri2012, LatifFerrara2016, 
Gallerani2017PASA, Mayer2019}.

In particular, two among the several facets which are highly debated and still need to be addressed 
concern the initial seeds of these SMBHs and their maximum accretion rate. 
As for the initial seeds of SMBHs, the three most popular formation scenarios \citep[for a review see][]{LatifFerrara2016}  are: 
{\sl{(i)}} the core-collapse of massive, Pop~III stars; 
{\sl{(ii)}} the collapse of the innermost region of a dense star cluster; and 
{\sl{(iii)}} the direct collapse BH channel 
\citep[e.g.][]{Bromm2003, Begelman2006, Mayer2010, Ferrara14, Pacucci15, Maio2018}. 
The aformentioned scenarios are still debated, and the likelihood of each of them is deeply connected 
(among other factors) to the redshift at which SMBH seeds formed and to the timescale over which they 
accreted gas to reach the mass they have at $z \sim 6$. Gas accretion should proceed at a fast pace, 
with BH accretion rates close to the Eddington accretion rate for long periods so as to let BH seeds 
reach their final mass by $z \sim 6$. In addition, the possibility of short-lived and intermittent episodes 
of super-critical accretion rate -- where the BH accretion rate overshoots the Eddington limit -- has been suggested to 
reconcile theoretical predictions with observations \citep[][]{Madau2014, Volonteri2015, Inayoshi2016, Regan2019}. 
Besides gas accretion, the concurrent channel for BH growth is merging with other BHs. 

Moreover, the problem of how BHs grow supermassive in the early universe is deeply intertwined both 
with the assessment of the contribution from AGN to the reionization of the universe 
\citep{Volonteri2009, Giallongo2015, Onoue2017, Hassan2018, Meyer2019, Trebitsch2020}, 
and with the early stages of BH-galaxy co-evolution \citep[e.g.][]{Lamastra2010, Merloni2010, Sarria2010, 
Willott2010, Portinari2012, Bongiorno2014, Valiante2014, Volonteri2016, Pensabene2020}.

High-redshift galaxies are complex ecosystems: 
they indeed have a multiphase interstellar medium (ISM), 
with gas spanning a wide range of temperatures, densities, and ionization states 
\citep[e.g.][for reviews, and references therein]{Wolfe2005, Carilli2013, Dayal2018}. 
As for $z \geq 6$ quasar-host galaxies, observations show that they 
typically have 
dynamical (gas and stellar) masses in the range $\sim 10^{10} - 10^{11}$~M$_{\odot}$, 
star formation rates (SFRs) from few hundreds to few thousands~M$_{\odot}$~yr$^{-1}$ 
\citep{Maiolino2005, Solomon2005, Wang2016, Trakhtenbrot2017a, 
Willott2017, Decarli2018, Venemans2018ApJ866, Bischetti2019, Shao2019}, 
molecular gas 
\citep[$> 10^{10}$~M$_{\odot}$][]{Walter2004, Shields2006, Venemans2017_845, Combes2018, Feruglio2018, LiJianan2020}, 
dust 
\citep[$> 10^8$~M$_{\odot}$; e.g.][]{Wang2008, Venemans2016, Carniani2019}, 
and outflows 
\citep[e.g.][]{Maiolino2012, Cicone2015, Bischetti2019, Stanley2019}. 
The cold and molecular gas phases play a key role, as they provide the reservoir of gas 
which fuels star formation (SF). 
The tight correlation between the SFR and the stellar mass (i.e. the main-sequence) is 
also already established at redshift $z \geq 6$ \citep{Bouwens2012, Salmon2015}: the normalization of this 
relation is observed to be higher than in the lower-redshift universe, thus implying higher SFRs and shorter 
gas depletion timescales for distant galaxies \citep{Solomon2005, Daddi2010}. 

Another piece of evidence which adds complexity to this picture is the role of stellar and quasar\footnote{The terms 
	quasar feedback and AGN feedback are often used interchangeably in this work.} feedback. 
Feedback is the complex set of processes by which SMBHs and supernovae (SNe) affect 
the evolution of their host galaxy and surrounding environment, and mainly develops through 
the injection of energy and momentum. 
These processes control structure formation and evolution across cosmic time: 
for instance, they shape galaxy morphology, affect the properties of the ISM, 
and regulate (or even quench) SF in galaxies 
\citep[e.g.][]{mcnamara2007, fabian2012, KormendyHo2013}. 
Feedback mechanisms can indeed prevent gas from being accreted or from effectively cooling (preventive feedback); 
they can remove gas from the innermost regions of forming structures where SF occurs (ejective feedback), 
or can suppress the SF efficiency 
(mainly via ISM heating and turbulence, e.g. \citealt{Alatalo2015, Costa2018}; 
but see \citealt{Bischetti2020} for different evidence). 
Feedback processes play a key role in determining the stellar-to-halo mass fraction and 
reducing the baryon coversion efficiency 
\citep[e.g.][]{Guo2010, Moster2010, Behroozi2013, Genzel2015, Pillepich2018_475, Bluck2020}. 
As a direct consequence, the low- and high-mass end of the galaxy stellar mass function is lowered and the shape predicted by theoretical models and simulations better agree with observations 
\citep{Croton2006, Puchwein2013, Hirschmann2014, Vogelsberger2014}.
Despite of the fact that one process can dominate over the others 
depending on the system properties and on cosmic time \citep[e.g.][]{ForsterSchreiber2019}, 
feedback mechanisms often occur in a simultaneous way, and it is hard to distinguish their imprints. 

Quasar feedback is expected to be the most important mechanism to suppress SF in massive systems, 
and AGN-driven outflows represent one of the main signatures of ongoing AGN activity. 
Other processes can contribute to suppress SF, e.g. stellar feedback, morphological quenching or gravitational heating, but it is unlikely that these processes alone (i.e. without the inclusion of quasar feedback) 
can keep massive systems quiescent. On the other hand, stellar feedback and environmental processes 
play the main role in regulating the star formation history (SFH) of lower-mass systems ($\leq 10^{11}$~M$_{\odot}$). 
Moreover, quasar feedback is fundamental to control the BH growth and the AGN activity itself, by regulating the evolution of physical properties of the gas surrounding the BH, and thus of BH accretion and luminosity.
However, it is still debated whether quasar feedback is the main driver of galaxy evolution and 
to what level it impacts on the physical properties of the bulk of the gas in galaxies. 
This is due to poor statistics and little availability of observations, especially at high redshift, 
where the SMBH activity or feedback is caught in the act \citep{Veilleux2005, fabian2012, Fiore2017}. 

The challenge of exploring the assembly of high-redshift systems 
and reproducing the growth of their SMBHs in the early universe 
can be tackled through cosmological simulations, which represent a unique theoretical tool 
\citep[e.g., among those focussing on the high-z universe,][]{Dubois2012, Bellovary2013, Costa2014MNRAS439, 
Feng2014, Feng2016, Fiacconi2017, Olsen2017, Barai2018, Lupi2019, Trebitsch2020Ob}. 
Simulations indeed allow to go through and connect subsequent evolutionary stages, 
rather than (observing) a single frame. 
Several numerical works have investigated interesting properties of high-redshift quasar-host galaxies 
in terms of e.g. 
their environment \citep{Costa2014MNRAS439}, 
the impact of quasar outflows on the host galaxy \citep{Barai2018}, 
and how cold flows from the large scale contribute to the growth of SMBHs \citep{Feng2014}. 
However, even if there is a general consensus on the key role played by SMBHs in the evolution of their 
host galaxy, details on the relative contribution of different processes in establishing final properties of systems 
are still debated. 
Moreover, predictions from simulations appear to be extremely sensitive to the different 
sub-resolution models adopted. 
As for the modelling of quasar feedback, for instance, models adopting kinetic injection of feedback energy 
\citep[e.g.][]{Barai2016, Barai2018} usually produce stronger feedback effects than those resulting from 
simulations assuming that feedback energy is either deposited as thermal only, or provided in a hybrid way. 
This difference mainly stems from the facts that kinetic energy is not radiated away when velocity is imparted to gas,
and that in this case thermalisation happens by construction later and at larger scales
\citep[see e.g. discussion in][and references therein]{Costa2020}. 
Cosmological simulations are undoubtedly crucial to test different scenarios and 
to assess the impact of different processes, as they allow 
the possibility of switching on and off different physical modules in the adopted code. 

Another appealing feature of cosmological simulations as for the investigation of how SMBHs reach the observed mass at $z=6$ is that they allow to explore the relative contribution of gas accretion and BH-BH merging to 
BH growth, and to quantify the impact of different assumptions for SMBH seed mass. 
Interestingly, for instance, \citet{Huang2019} showed that the adopted value of the BH seed mass 
shapes the BH early growth and merging history, even if the final mass of the SMBH at $z=6$ is not sensitive 
to the assumed seed mass. 
Unfortunately, state-of-the-art cosmological simulations to date usually adopt rather crude seeding prescriptions 
for BHs, as they seed BH particles with a mass that is assumed to be already the rusult of the formation of 
SMBH seeds, whose details are not captured. 

An additional task that cosmological simulations can accomplish is to shed some light on the progenitors of SMBHs observed at redshift $z \sim 6$. Indeed, to spot BHs as massive as $\sim 10^7$~M$_{\odot}$ in the distant universe 
has proven less successful so far: this questions if and why they are so rare, and whether they are possibly 
obscured or too faint to be detected by current facilities. 
This theoretical approach has also a twofold, paramount importance: results of simulations can be indeed used 
to interpret observational results and to guide future surveys. 

The framework that we have just outlined opens to new challenging tasks. 
The goal of this paper is to: (a) investigate how BHs grow supermassive in the early universe, 
and (b) explore their impact on the host galaxy and surrounding ISM. 
To this aim, we introduce a new suite of cosmological hydrodinamical simulations, 
where we take advantage of a detailed modelling of both the ISM physics, and 
BH accretion and feedback.  
The main questions that we aim at addressing with this introductory work are the following: 
what is the impact of stellar and quasar feedback on the ISM of high-redshift ($z \simeq 6$) galaxies hosting SMBHs? 
How does thermal quasar feedback affect the growth history of SMBHs in the early universe? 
Do SMBH properties correlate with those of the hosting galaxies? 
What is the relative contribution of stellar and quasar feedback in promoting outflows?
Can we constrain galactic outflow features at high redshift?

The outline of this Paper is as follows. 
Section~\ref{sec:sims} introduces the set of cosmological simulations and 
describes the main features of the sub-resolution model adopted. 
In Sections~\ref{sec:results}~and~\ref{sec:discussion}, we present and discuss our results. 
We draw conclusions in Section~\ref{sec:conclusions}.

\section{The suite of cosmological simulations} 
\label{sec:sims}

In this Section, we describe the initial conditions (Section~\ref{sec:ICs}) of our cosmological simulations,
the sub-resolution model that we adopt (Section~\ref{sec:MUPPI}), and the set of simulations that we 
performed (Section~\ref{sec:simset}). 
Simulations have been carried out with the TreePM+SPH (smoothed particle hydrodynamics) \textsc{GADGET3} 
code, a non-public evolution of the \textsc{GADGET2} code \citep{Springel2005}. 
We adopt the advanced formulation of SPH presented in \citet{beck2015} 
and introduced in cosmological simulations adopting our sub-resolution model by \citet{Valentini2017}. 
This improved formulation of SPH features a high-ordel kernel function, an artificial conduction term, 
a correction for the artificial viscosity, and a wake-up scheme, among the main refinements.

\subsection{Initial conditions} 
\label{sec:ICs}

We used the \textsc{MUSIC}\footnote{\url{https://www-n.oca.eu/ohahn/MUSIC/}} 
software \citep{Music2011} to generate the initial conditions (ICs).
The assumed $\Lambda$CDM cosmology is the following: $\Omega_{\rm m}=0.3089$, 
$\Omega_{\rm \Lambda}=0.6911$, $\Omega_{\rm baryon}=0.0486$, $\sigma_8 = 0.8159$, $n_s=0.9667$, 
and $H_{\rm 0}=100 \,h$~km~s$^{-1}$~Mpc$^{-1}=67.74$~km~s$^{-1}$~Mpc$^{-1}$. 
These parameters are in agreement with the results of \citet{Planck2016}.  
First, a parent, DM-only simulation of a cubic volume of size $L=148$~cMpc (i.e. comoving Mpc)\footnote{We 
	use the letter {\sl c} before the corresponding unit to refer to {\sl comoving} distances (e.g. ckpc), 
	while by analogy the letter {\sl p} stands for {\sl physical} units (e.g. pkpc). 
	When not explicitly stated, we are referring to physical distances.} 
is run starting at $z=100$ down to $z=6$, with periodic boundary conditions. 
We used $512^3$ DM particles, the resulting mass resolution being $9.4 \cdot 10^8$~M$_{\odot}$. 
The gravitational softening length is set to $5.8$~ckpc (corresponding to $1/50$ of the mean inter-particle spacing).
At $z=6$, subhaloes have been identified by means of the \textsc{SUBFIND} algorithm 
\citep{Springel2001subfind, Dolag2009}. 
The subhalo finder algorithm provides the mass of each subhalo, as well as the coordinates of its centre, 
which is represented by the position of the subhalo particle with the minimum value of the gravitational potential.
The subhalo mass is used to compute the virial radius of the structure. 
The virial radius is defined as the radius that encloses an overdensity of $\Delta_{\rm vir}(z)$ times the
critical density of the universe at that redshift \citep{Barkana2001}, where $\Delta_{\rm vir}(z)$ is the overdensity of 
virialised structures in a flat universe \citep{Bryan1998}.
The main properties of subhaloes, along with their distribution in the parent, DM-only simulation at $z=6$ 
are discussed in Appendix~\ref{DMsim}. 

We chose a DM subhalo and re-simulated it with a higher resolution, zoomed-in simulation. 
Our target DM subhalo has been selected among those subhaloes which are at least as massive 
as $10^{12}$~M$_{\odot}$ at $z=6$, so as to be the eligible halo of a quasar-host galaxy 
(see Appendix~\ref{DMsim} for further details). 
The target subhalo has a mass of $1.12 \cdot 10^{12}$~M$_{\odot}$ and a virial radius of $r_{\rm vir, DM}=48.1$~pkpc. 
We identified the DM particles within $2.5 \, r_{\rm vir, DM}$ at $z=6$ and traced them back to their initial 
positions at $z=100$. The positions occupied by these particles define a Lagrangian region of $144.4$~(cMpc)$^3$. 
By approximating this volume with a cube, the side of the zoom-in volume is $5.25$~cMpc. 
We increased the resolution of the ICs by adding three more levels of refinement within the 
Lagrangian region with the \textsc{MUSIC} software, and included baryons. 
In the final zoomed-in simulation, 
the highest resolution DM particles have a mass of $m_{\rm DM} = 1.55 \cdot 10^6$~M$_{\odot}$,
while gas particles have $m_{\rm gas} = 2.89 \cdot 10^5$~M$_{\odot}$. 
Softening lengths for high-resolution DM and baryonic particles are as follows: 
$\epsilon_{\rm DM} = 0.72$~ckpc and $\epsilon_{\rm bar} = 0.41$~ckpc, respectively, which translate 
in a force resolution 
of $\epsilon_{\rm DM} = 103$~ppc and $\epsilon_{\rm bar} = 59$~ppc (i.e. physical pc at $z=6$). 

The ICs of the zoomed-in simulation have $\sim 2 \times 3.14 \cdot 10^6$ particles 
of gas and high-resolution DM, and $\sim 1.35 \cdot 10^8$ lower-resolution DM particles. 
We verified with a DM-only test run whose resolution is analogous to that of the zoomed-in simulation that 
the main subhalo at $z=6$ is not contaminated by lower-resolution DM particles. 
There are no lower-resolution DM particles within the virial radius of the main progenitor of the target subhalo 
at all redshifts; moreover, the fraction of contaminating particles within twice the virial radius of the main progenitor 
of the target subhalo is below $1$\% in mass and $0.1$\% in number.
The ICs of the zoomed-in simulation are evolved until $z=6$ with different flavours of baryonic physics: 
results will be discussed in Section~\ref{sec:results}.

\subsection{The sub-resolution model} 
\label{sec:MUPPI}

The sub-resolution model that we adopt for our cosmological simulations is \textsc{MUPPI} 
(MUlti Phase Particle Integrator): 
it represents a multiphase ISM and accounts for a variety of physical processes which occur on scales not explicitly resolved.
The model has been introduced and thoroughly described in \citet{muppi2010, muppi2014} and 
\citet{Valentini2017, Valentini2019, Valentini2020}: in this Section, we outline its main features, while 
we refer the reader to the aformentioned papers for a more comprehensive discussion and any further details.

\subsubsection{Star formation and stellar feedback} 
\label{sec:CSF}

Our sub-resolution model describes a multiphase ISM. The multiphase particle represents its essential element: 
it is made up of hot and cold gas in pressure equilibrium, and a possible further stellar component. 
A gas particle enters a multiphase stage should its density rise above a 
density threshold ($n_{\rm H, \, thres}=0.01$~cm$^{-3}$)
and its temperature decrease below a 
temperature threshold ($T_{\rm thresh}=5 \cdot 10^4$~K). 

We adopt a set of ordinary differential equations to describe mass and energy flows among different components:  
radiative cooling makes hot gas condense into a cold phase (whose temperature is fixed to $T_{\rm c}=300$~K), 
while some cold gas evaporates due to the destruction of molecular clouds. 
We rely on an $H_2$-based SF law to compute the instantaneous SFR of each 
multiphase particle. 
A fraction $f_{\rm mol}$ of the cold gas mass $M_{\rm c}$ is in the molecular phase: the molecular 
gas is then converted into stars over a timescale which is the dynamical time of the cold gas ($t_{\rm dyn, c}$), and 
according to an efficiency ($f_{\star} = 0.06$) aimed at capturing the average SF efficiency. 
Hence, the SFR associated to a multiphase particle reads:
$\,\, \dot{M}_{\rm sf} \:=\: f_{\star} \, f_{\rm mol} \, M_{\rm c} / t_{\rm dyn} \,$. 
The molecular fraction $f_{\rm mol}$ is computed according to the phenomenological prescription by \citet{blitz2006}:
$\, f_{\rm mol} \:=\: 1/(1+P_0/P) \,\,$,
where $P$ is the hydrodynamic pressure of the gas particle and 
$P_0$ is the pressure of the ISM at which $f_{\rm mol}=0.5$, this parameter being derived from observations 
(we adopt a constant value $P_0 / {\text {k}}_{\rm B}= 2 \cdot 10^4$~K~cm$^{-3}$)\footnote{The 
	effective density threshold for the SF is $n_{\rm thresh, sf} \simeq 66.7$~cm$^{-3}$. 
	Equation~$\, f_{\rm mol} \:=\: 1/(1+P_0/P) \,\,$ implies that 
	{$\, n_{\rm thresh, sf} \, T_{\rm c} = 2 \cdot 10^4$~K~cm$^{-3}$}, 
	assuming $n_{\rm thresh, sf}$ as the number density of the cold gas for which $f_{\rm mol}=0.5$, 
	and plugging in the adopted value for $P_0$. 
	On the other hand, $n_{\rm thres}$ is the density threshold 
	to let a gas particle sample the multiphase ISM.}. 
SF is implemented according to the stochastic model introduced by \citet{SpringelHernquist2003}, 
to spawn stellar particles from gas particles. 

Energy contributed by SN explosions counterbalances radiative cooling, 
along with the hydrodynamical term accounting for
shocks and heating (cooling) due to gravitational compression (expansion) of gas.
Stellar feedback releases energy both in thermal and kinetic forms. 
The thermal stellar feedback energy 
$\,\, \Delta E_{\rm fb, therm}= f_{\rm fb, therm} \, E_{\rm SN} \, \Delta M_{\star} / M_{\star, \rm SN}\,\,$ 
is supplied by each multiphase star-forming particle to neighbours 
within a cone (whose half-opening angle is $\vartheta=30^{\circ}$), in a given time-step. 
Here, $f_{\rm fb, therm}$ describes the thermal stellar feedback efficiency (i.e. the fraction of 
$E_{\rm SN} = 10^{51}$~erg which is actually coupled to the ISM), 
$M_{\star, \rm SN}$ is the stellar mass that is required on average to have a single SN~II, 
and $\Delta M_{\star}$ represents the mass of the multiphase particle that has been converted into stars. 
As for kinetic stellar feedback, which is a key process to drive galactic outflows, 
our sub-resolution model adopts the galactic outflow model introduced in \citet{Valentini2017}. 
According to this model, the ISM is isotropically provided with kinetic stellar feedback energy. 
Each star-forming particle supplies the energy: 
$\,\, \Delta E_{\rm fb, kin}= f_{\rm fb, kin} \, E_{\rm SN} \, \Delta M_{\star} / M_{\star, \rm SN}\,\,$ 
isotropically, to all the wind particles\footnote{Wind particles are gas particles which sample galactic outflows 
	and are hydrodynamically decoupled from the surrounding medium for a lapse of time 
	\citep[see][for details]{Valentini2020}.} 
within the smoothing length, with kernel-weighted contributions. 
Here, $f_{\rm fb, kin}$ is the kinetic stellar feedback efficiency (see Section~\ref{sec:simset} for adopted values). 
Wind particles receiving energy use it to increase their velocity along their least resistance
path, since they are kicked against their own density gradient. 
We refer the interested reader to the aforementioned papers 
\citep{muppi2010, muppi2014, Valentini2017, Valentini2019} for a thorough description of 
all the processes included in our sub-resolution model; in particular, Section~$2$ of \citet{Valentini2020} provides 
a comprehensive summary and can be taken as a reference for the values of all the parameters not explicitly 
mentioned here. 
A gas particle exits the multiphase stage when its density drops
below a density threshold ($0.2 \rho_{\rm thresh}$) or once a maximum time 
(set by the dynamical time of the cold gas) elapses. 

MUPPI also features chemical evolution and enrichment, these processes being self-consistently 
accounted for following the model by \citet{tornatore2007}. Star particles (each describes a simple 
stellar population, assuming a \citealt[][]{ChabrierIMF2003} initial mass function) produce and release 
heavy elements, which are distributed to neighbouring gas particles 
with kernel-weighted contributions. 
We follow the production of heavy elements ($13$ different metals, plus Hydrogen and Helium) 
released by aging and exploding stars adopting sets of stellar yields. 
We assume the stellar yields provided by \citet{Thielemann2003} for SNe~Ia and 
the mass- and metallicity-dependent yields by \citet{Karakas2010} for intermediate and low-mass stars 
during the AGB phase. As for SNe~II, we use the mass- and metallicity-dependent yields 
by \citet{WoosleyW1995} and \citet{Romano2010}, as detailed in \citet{Valentini2019}.
Each element separately contributes to the gas cooling rate, that is modelled according to \cite{wiersma2009}. 
To infer cooling rates, the effect of a spatially uniform, time-dependent ionizing cosmic background 
\citep{HaardtMadau2001} is accounted for.

\subsubsection{BHs and quasar feedback} 
\label{sec:AGN}

BHs are treated as collisionless sink particles which are seeded in DM haloes whose mass 
is larger than 
$M_{\rm DM, seed} = 1.48  \cdot 10^9$~M$_{\odot}$ (see below), and which do not already host a BH. 
A Friends-of-Friends (FoF) algorithm, run on-the-fly, identifies 
DM haloes. 
BHs are first introduced with a seed mass 
$M_{\rm BH, \, seed}=1.48 \cdot 10^5$~M$_{\odot}$ 
\citep[similar values for $M_{\rm DM, seed}$ and $M_{\rm BH, \, seed}$ have been previously 
adopted by e.g.][]{Costa2014MNRAS439, Barai2018}, in the position of the minimum potential of the halo. 
Seeding presciptions like the one we use are meant to capture the result of the formation of 
direct collapse BHs (see Section~\ref{sec:introduction}). 
BHs are pinned to the minimum of the gravitational potential, to prevent them from wandering from 
the centre of the halo in which they reside because of numerical artefacts \citep[][]{Wurster2013}. 
Hence, at each time-step we shift the BH towards the position of the particle with the absolute 
minimum value of the local gravitational potential within the gravitational softening of the BH, if not already there 
\citep[as also done by e.g.][]{RagoneFigueroa2013, Schaye2015, Weinberger2017, Pillepich2018}. 
 
Once seeded, BHs grow as a result of two processes: gas accretion and mergers with other BHs. 
We model gas accretion onto the BH via the Bondi-Hoyle-Lyttleton accretion 
solution \citep{Hoyle1939, Bondi1944, Bondi1952}. 
The Bondi-like accretion rate is numerically estimated as:
\begin{equation}    
\dot{M}_{\rm B} = \frac{4 \, \pi \, G^2 \, M_{\rm BH}^2 \, \langle \rho \rangle}{(\langle c_{\rm s}\rangle^2 + \langle v\rangle^2)^{3/2}}  \,\,\,,
\label{BondiMdotAve}
\end{equation}
where $G$ is the gravitational constant \citep{Springel2005e361}. In equation~(\ref{BondiMdotAve}), 
the density of the gas $\langle \rho \rangle$, its sound speed $\langle c_{\rm s}\rangle$, and the velocity $\langle v\rangle$ of the BH relative to the gas are 
calculated by averaging over SPH quantities of the gas particles within the BH smoothing length, with 
kernel-weighted contributions. 
We refer to the smoothing length of a BH particle by analogy with gas particles, as the radius $h_{\rm i}$ 
of the sphere centred on the considered particle, which contains a given number of neighbour particles 
\citep[$\sim 200$, for the kernel function that we adopt; see][for details]{beck2015}. 
In our simulations, we distinguish between hot and cold gas accretion \citep[see][]{Valentini2020}. 
We assume a temperature threshold 
$T_{\rm split} = 5 \cdot 10^5 $~K \citep[e.g.][]{Steinborn2015} to differentiate hot from cold gas. 
The accretion rates for the hot and cold phases are estimated separately according to 
equation~(\ref{BondiMdotAve}). Once $\dot{M}_{\rm B, \, h}$ and $\dot{M}_{\rm B, \, c}$ are retrieved, 
the gas accretion rate $\dot{M}_{\rm accr}$ is given by the sum of both hot and cold gas accretion, 
and is limited to the Eddington accretion rate, i.e.:
\begin{equation}    
\dot{M}_{\rm accr} = {\text{min}}  (\dot{M}_{\rm B, \, h} + \dot{M}_{\rm B, \, c}, \dot{M}_{\rm Edd}) \,\,\,.
\label{Mdot_limited}
\end{equation}
As a result of the gas accretion process, the BH can absorb neighbour gas particles according to the 
stochastic scheme by \citet{Springel2005e361}. 
We do not assume any boost factor in equation~(\ref{BondiMdotAve}), neither for the hot nor 
for the cold gas accretion in our simulations.  
Indeed, our resolution and sub-resolution modelling of the ISM physics allow us to achieve 
a fair description of the accretion process and final BH masses in line with observations 
without the need for a boost factor \citep[see also Section~\ref{sec:discussion},  and the discussion 
in Sections~$1$~and~$3.2$ of][for further details and caveats]{Valentini2020}.  

Moreover, we further improve the modelling of gas accretion by taking into account the angular momentum 
of the accreting gas \citep[see also][]{RosasGuevara2015, Schaye2015}. 
We limit the inflow of the cold gas which has a high angular momentum: gas with rotational support 
is indeed expected to depart significantly from the Bondi assumptions, and prevented from being directly accreted. 
Therefore, the contribution to the gas accretion rate from the cold gas (entering in 
equation~(\ref{Mdot_limited})) is: 
\begin{equation}    
\dot{M}_{\rm B, \, c} = \dot{M}_{\rm B, \, c} \, {\text{min}}  (1, \mathcal{L}_{\rm AM}) \,\,\,, 
\label{Mdot_limited_mango}
\end{equation}
where $\mathcal{L}_{\rm AM}$ is the gas accretion rate limiter, i.e.: 
\begin{equation}    
\mathcal{L}_{\rm AM} = \frac{1}{C_{\rm visc}} \, \biggl( \frac{c_{\rm s, \, c}}{V_{\phi}} \biggr)^3   \,\,\,.
\label{AngMomLimiter}
\end{equation}
In equation~(\ref{AngMomLimiter}), $C_{\rm visc} = 2 \pi$ is a constant parameter aimed at capturing 
the viscosity of the accretion disc at the sub-resolution level, 
$c_{\rm s, \, c}$ is the sound speed of the cold 
($T < T_{\rm split}$) gas, 
and $V_{\phi}$ is the rotational velocity of the cold gas surrounding the BH \citep[i.e. within its smoothing length; see][for details]{Valentini2020}. 
We refer to \citet{Valentini2020} for a thorough investigation into the impact 
of the angular momentum dependent gas accretion. 

The actual mass growth rate of the BH reads: 
\begin{equation}    
\dot{M}_{\rm BH} = (1 - \epsilon_{\rm r}) \, \dot{M}_{\rm accr} \,\,\,. 
\label{MBH_actual}
\end{equation}
A small fraction $\epsilon_{\rm r} \, \dot{M}_{\rm accr}$ is radiated away. 
In the quasar feedback process which ensues from gas accretion, the BH bolometric luminosity thus reads: 
\begin{equation}    
L_{\rm r} = \epsilon_{\rm r} \, \dot{M}_{\rm accr} \, c^2  = \frac{\epsilon_{\rm r}}{1 - \epsilon_{\rm r}} \, \dot{M}_{\rm BH}  \, c^2   \,\,\,.
\label{Luminosity}
\end{equation}
As for the radiative efficiency, we assume $\epsilon_{\rm r}=0.03$\footnote{Such a value for the radiative efficiency 
	is compatible within a factor of~$2$ with the minimum value of the accretion efficiency for an accretion disc 
	surrounding a non-spinning BH \citep{Shakura1973, Novikov1973}, and is in agreement with results of 
	\citet{Sadowski2017, Trakhtenbrot2017}. 
	When low values of $\epsilon_{\rm r}$ (i.e. non-spinning BHs) are assumed, 
	so that $\epsilon_{\rm r}/(1 - \epsilon_{\rm r}) \approx  \epsilon_{\rm r}$ in eq.~(\ref{Luminosity}), 
	final results are mainly sensitive to the product  
	$\epsilon_{\rm r} \, \epsilon_{\rm f}$ (see eq.~(\ref{Luminosity2})), although 
	a low radiative efficiency always helps the black hole grow faster (see eq.~(\ref{MBH_actual})).}
A tiny part of the radiated luminosity $L_{\rm r}$ is then coupled to the ISM as AGN feedback energy, 
the feedback energy per unit time being:
\begin{equation}    
\dot{E}^{\rm AGN}_{\rm fb, \rm tot} = \epsilon_{\rm f} \, L_{\rm r} \approx \epsilon_{\rm f} \, \epsilon_{\rm r} \,  \dot{M}_{\rm BH}  \, c^2 \,\,\,, 
\label{Luminosity2}
\end{equation}
where $\epsilon_{\rm f}$ is the feedback efficiency. 
We adopt $\epsilon_{\rm f} = 10^{-4}$: we tuned this efficiency in order to match the normalisation of the 
black hole to stellar mass relation ($M_{\rm BH} - M_{\star}$, see Section~\ref{BH_properties}) 
at $z=6$\footnote{The value which $\epsilon_{\rm f}$ is set to in our simulations is lower than commonly 
	adopted values by a factor of~$\sim 10$. 
	Values assumed in other works span the range~$\sim 10^{-4} - 0.15$ 
	\citep[e.g.][]{Ostriker2010, Dubois2012, Costa2014MNRAS439, Lupi2019, Trebitsch2020}. 
	Note that sometimes such higher values are calibrated in order to reproduce the normalization of the 
	$M_{\rm BH} - M_{\star}$ relation in the local universe \citep[e.g.][]{Dubois2012, Costa2014MNRAS439}, which 
	is below that inferred from observation at $z=6$ by a factor of~$\sim15$ \citep{Wang2010}. 
	When our model is used to reproduce the $M_{\rm BH} - M_{\star}$ relation at $z=0$, a value 
	$\epsilon_{\rm f} = 0.01$ is adopted, as discussed in \citet{Valentini2020}.} \citep{Wang2010, Pensabene2020}.
This quasar feedback energy is coupled thermally and isotropically to the BH neighbouring gas particles. 

The quasar feedback energy is distributed to all gas particles 
within the smoothing sphere of the BH, in a kernel-weighted fashion. 
Quasar feedback energy contributions assigned to single-phase particles within the BH smoothing volume 
increase their temperature and specific internal energy. 
Quasar feedback energy assigned to the multiphase particles within the BH kernel (the vast majority) is distributed 
to both their hot and cold phases. The covering factors of the hot and cold gas within each multiphase particle 
determine the fraction of feedback energy provided to the two phases. The physical idea behind this modelling 
assumes that the larger is the volume occupied by the cold gas, the larger is the amount of energy that it absorbs. 
The quasar feedback energy coupled to the hot gas within the multiphase particle is used to increase its temperature,  while the quasar feedback energy provided to the cold phase produces the evaporation of the cold gas, 
whose mass is brought to the hot phase. Further details about the modelling of the AGN feeding and feedback 
processes can be found in \citet{Valentini2020}.

\begin{figure*}
\newcommand{\captionfonts}{\small}
\begin{minipage}{\linewidth}
\centering
\includegraphics[trim=2.5cm 0.cm 2.5cm 0.cm, clip, angle=90, width=1.\textwidth]{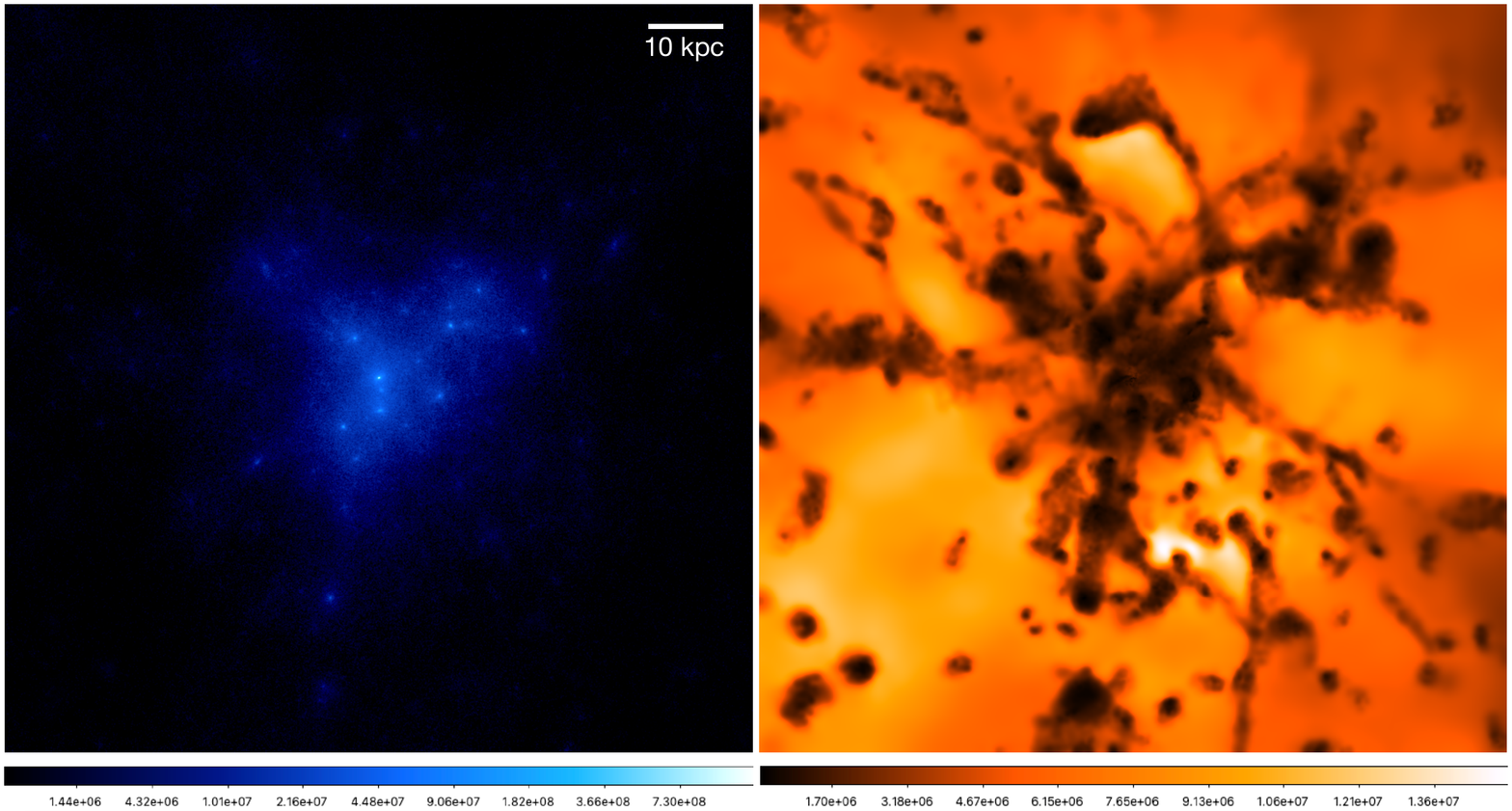} 
\end{minipage} 
\caption{{\sl Left panel: } Map of total (DM, stars, and gas) mass surface density (in units of g/cm$^2$) 
	of the reference simulation {\sl AGN\_~fid}. 
	{\sl Right: } Projected, smoothed gas temperature map, in units of K. 
	We show a box of $100$~pkpc a side at redshift $z = 6$, 
	the projection being performed along the $z$-axis (over $100$~pkpc). 
	Both the maps are centred on the centre of the most massive subhalo.}
\label{IntroMaps} 
\end{figure*}

The other process contributing to the BH growth is BH-BH merging: two BHs are merged should their distance 
be smaller than twice their gravitational softening length (the same as $\epsilon_{\rm bar}$), 
and if their relative velocity $v_{\rm BH - BH} < 0.5 \, \langle c_{\rm s}\rangle$. 
The resulting BH lies at the position of the most massive one between the two BHs that 
undergo merging.

\begin{figure*}
\newcommand{\captionfonts}{\small}
\begin{minipage}{\linewidth}
\centering
\includegraphics[trim=0.cm 0.cm 0.cm 0.cm, clip, width=1.\textwidth]{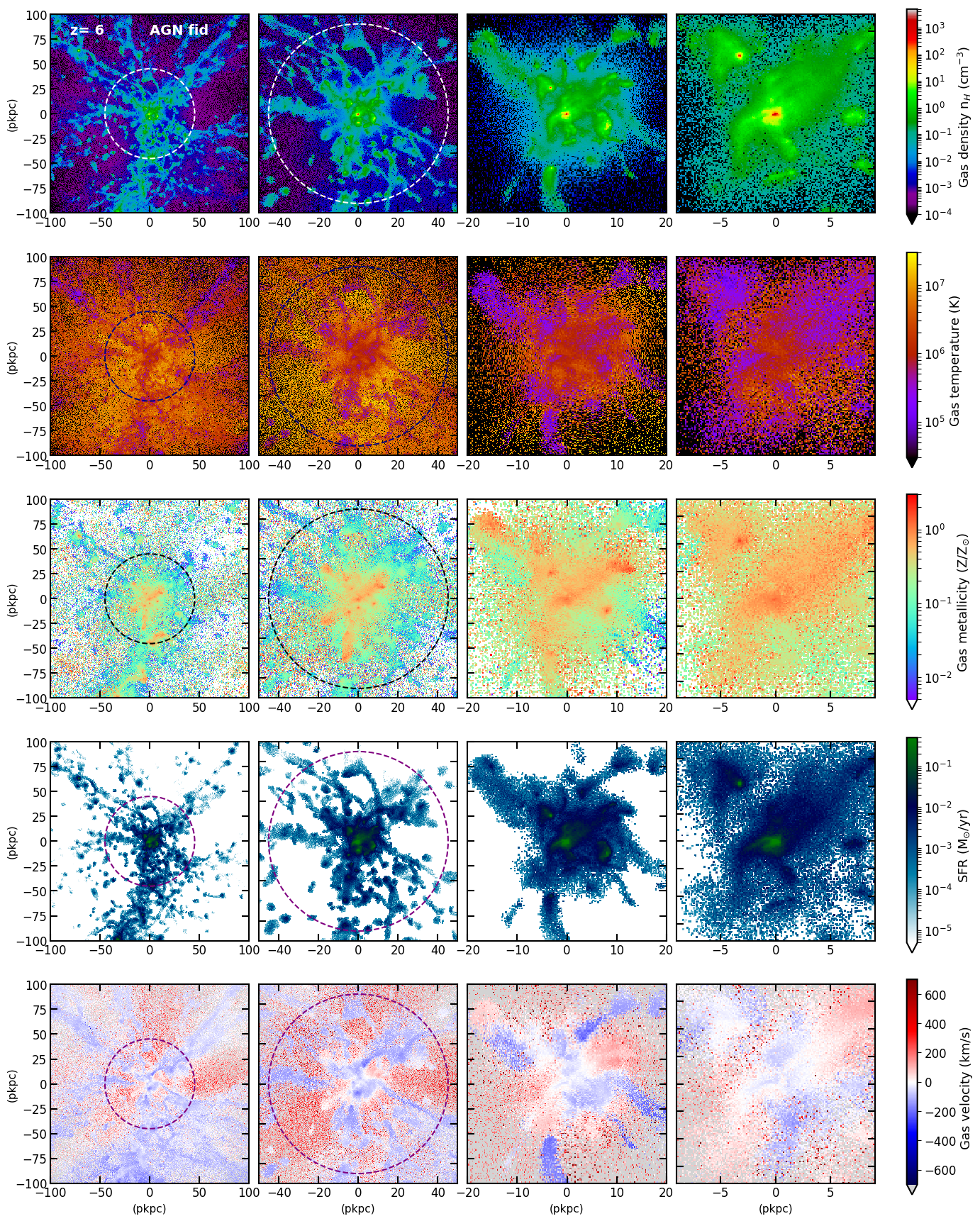} 
\end{minipage} 
\caption{Overview of the fiducial simulation {\sl AGN\_~fid}, at redshift $z = 6$. 
	We show gas density ({\sl first row}),
	gas temperature ({\sl second row}), 
	gas metallicity ({\sl third row}),
	the SFR of gas particles ({\sl fourth row}), 
	and the mass-weighted, radial velocity of gas particles ({\sl bottom row}). 
	We progressively zoom-in from left to right: 
	the {\sl first and second columns} show a box of $200$~pkpc and $100$~pkpc a side, respectively, 
	the projection being performed along the $z$-axis (over $200$~pkpc and $100$~pkpc, respectively). 
	The radius of the dashed circumference shows the virial radius 
	of the central, target halo (r$_{\rm vir}=45.17$~pkpc). 
	The {\sl third column} shows a box of $40$~pkpc (projection is over $20$~pkpc along the $z$-axis), 
	while in the {\sl fourth column} we consider a box of $18$~pkpc 
	(projection is over $9$~pkpc along the $z$-axis). 
	All the maps are centred on the centre of the most massive subhalo.}
\label{IntroMapsZoom} 
\end{figure*}

\begin{figure*}
\newcommand{\captionfonts}{\small}
\begin{minipage}{\linewidth}
\centering
\includegraphics[trim=0.cm 0.cm 0.cm 0.cm, clip, width=1.\textwidth]{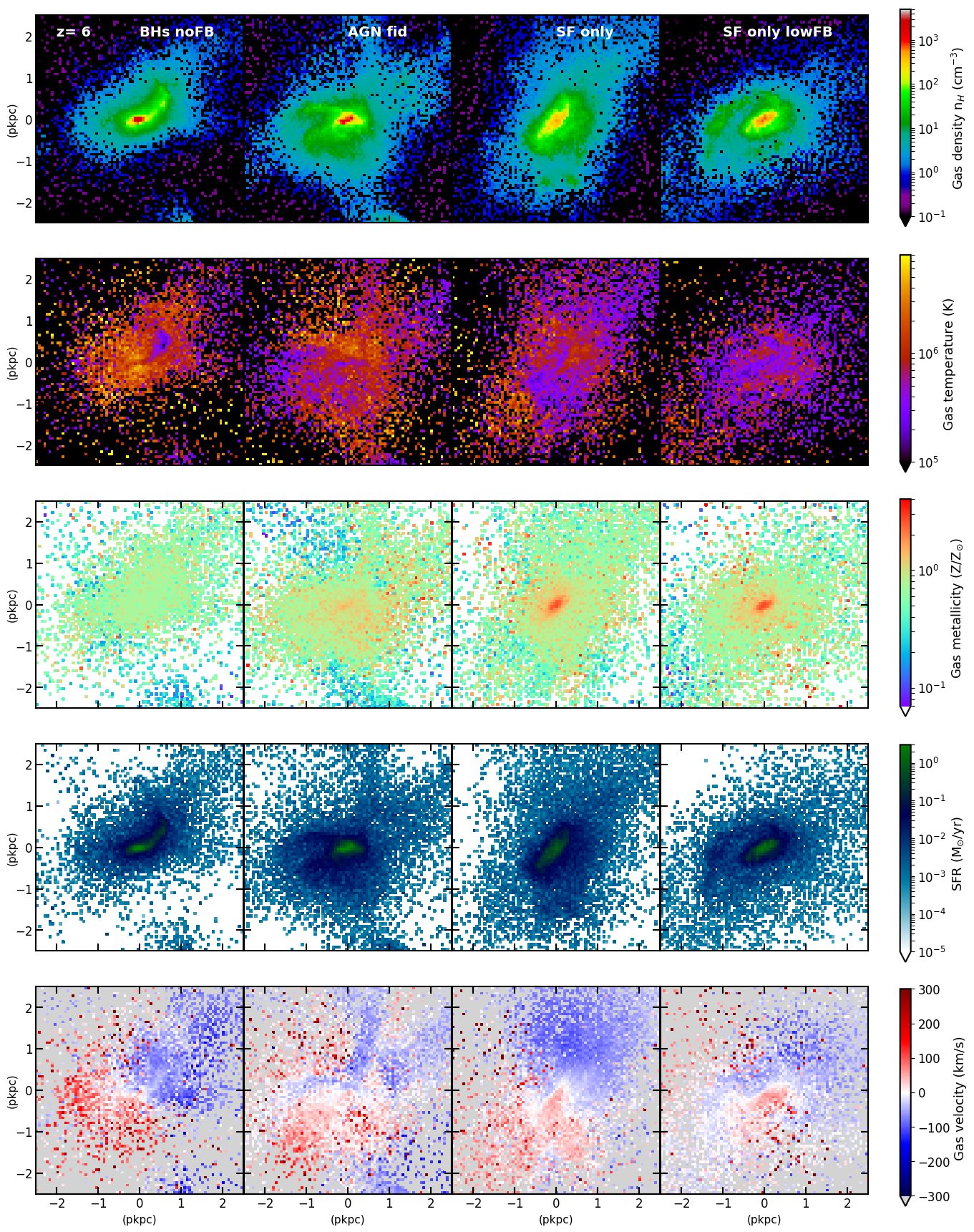} 
\end{minipage} 
\caption{Overview of the four simulated galaxies, at redshift $z = 6$. 
	We show gas density ({\sl first row}),
	gas temperature ({\sl second row}), 
	gas metallicity ({\sl third row}),
	the SFR of gas particles ({\sl fourth row}), 
	and the mass-weighted, radial velocity of gas particles ({\sl bottom row}). 
	Each column considers a different simulation: 
	the {\sl first column} shows the simulation {\sl BHs\_noFB}, 
	the {\sl second column} depicts {\sl AGN\_~fid};
	the {\sl third column} shows {\sl SF\_only}, and 
	the {\sl fourth column} {\sl SF\_only\_lowFB}. 
	Each box has a side of $5$~pkpc, 
	and quantities are averaged along the $z$-axis (over $1.2$~pkpc). 
	Bin size is comparable with the softening length of baryonic particles in the simulations. 
	All the maps are centred on the centre of the most massive subhalo.}
\label{IntroMaps4} 
\end{figure*}

\subsection{The set of simulations} 
\label{sec:simset}

In this work, we show and discuss the results of five different simulations that we carried out. 
They differ from one another as for the physical processes included or features in the sub-grid modelling.
Details are as follows: 

\begin{itemize}
\item {\sl AGN\_~fid} is the reference simulation with stellar and quasar feedback. 
				Being the properties of the central BH and its host galaxy 
				(i.e. BH mass and stellar mass of the quasar host; see Section~\ref{BH_properties}) 
				in agreement with the $M_{\rm BH} - M_{\star}$ relation at $z=6$, this is our fiducial model. 
				It includes 
				all the physical processes whose complex interplay we aim at investigating in this work. 
\item {\sl BHs\_noFB} is the simulation which includes BHs and the gas accretion process, but no quasar feedback; 
\item {\sl SF\_only} is the simulation where BHs are not included and the quasar feedback is not taken into account. 
It features SF and stellar feedback: the fiducial value $f_{\rm fb, kin} = 0.12$ 
for the stellar feedback efficiency is adopted here \citep[see][]{Valentini2018, Valentini2019},  
as well as in {\sl AGN\_~fid} and {\sl BHs\_noFB}; 
\item {\sl SF\_only\_lowFB} is a simulation where BHs and the ensuing feedback are not included, 
similar to {\sl SF\_only}. In this simulation, a lower kinetic stellar feedback efficiency is assumed with 
respect to {\sl SF\_only}, 
i.e. $f_{\rm fb, kin} = 0.05$ 
(the thermal stellar feedback efficiency is always set to $f_{\rm fb, therm} = 0.2$). 
Should a lower amount of stellar feedback energy be coupled to the ISM in the kinetic form,
the resulting outflows are triggered with a lower velocity.
\item {\sl AGN\_highFB} is a simulation analogous to the fiducial {\sl AGN\_~fid}, 
but adopts a feedback efficiency $\epsilon_{\rm f} = 10^{-3}$, 
i.e. higher than the reference value by a factor of $10$. 
The central, most massive BH mass of the quasar-host galaxy in this model is in agreement 
with the $M_{\rm BH} - M_{\star}$ relation in the local universe (whose normalization is 
observed to be lower than that at high redshift, see Section~\ref{BH_properties}). 
\end{itemize}

This suite of simulations has been designed with the aim to investigate:
{\sl i)} the relative impact of the SF and quasar feedback processes; 
{\sl ii)} the effect of the AGN feedback on the ISM of the galaxies hosting BHs, and on the growth of the BHs themselves; 
{\sl iii)} the possibility that stellar feedback alone is main driver of peculiar observational features, 
should galactic outflows be launched with different velocities.  

Besides the aforementioned simulations that we are going to consider in detail, we also performed 
several preliminary simulations (see Table~\ref{tab:otherSims} in Appendix~\ref{OtherSims}), 
which have been preparatory to this analysis. 
In particular, they have been fundamental to calibrate AGN feedback efficiencies 
and to explore the parameter space of our sub-resolution model when used to investigate the high-z universe.

\section{Results} 
\label{sec:results}

In this Section, we introduce our results. 
We will first show (Section~\ref{sims_overview}) a broad overview of four simulations among those 
presented in Section~\ref{sec:simset}. Then, we will focus in detail on the properties of 
the ISM of the quasar-host galaxy (Section~\ref{host_galaxies}), BHs (Section~\ref{BH_properties}), 
and inflow/outflow (Section~\ref{outflows_res}) in the reference simulation {\sl AGN\_~fid}.

\subsection{Overview of the simulations}
\label{sims_overview}

We start our analysis by investigating the final properties of the central galaxy in different simulations, at $z=6$. 
The central galaxy is the galaxy located in the most massive sub-halo, which hosts the most massive BH 
(should BHs be present in simulations {\sl AGN\_~fid} and {\sl BHs\_noFB}). 

Figure~\ref{IntroMaps} introduces our reference simulation {\sl AGN\_~fid}. 
We show the projected density of DM, stars, and gas (left panel) and 
the projected, smoothed, mass-weighted gas temperature map (right panel), at redshift $z = 6$. 
The Figure pictures the connection between the stellar (left panel) and gaseous (right panel) 
components in the centre of one of our simulated structures. 
We see clumps and filaments of denser and colder ($T \sim 10^5 - 10^6$~K, see also Figure~\ref{IntroMapsZoom}) 
gas, which is mainly inflowing and feeding the central galaxy, surrounded by a hotter and more diffuse phase. 

Figure~\ref{IntroMapsZoom} shows a progressive zoom-in on the central galaxy 
of the reference simulation {\sl AGN\_~fid}. 
We analyse gas density, temperature, gas metallicity, 
the SFR of gas particles, 
and the mass-weighted, radial velocity of gas particles with respect to the centre of the target sub-halo.  
Colours encode mean SPH quantities for gas particles in each spatial bin for all the panels but for those 
where the SFR is analysed. In this latter case, we consider the sum of the SFR of each gas particle in the bin, 
to account for the total SFR contributed by the star-forming gas in the bin. 
As for gas temperature, we consider the SPH estimate for single-phase particle and 
the mass-weighted average of hot and cold gas temperatures for multiphase particles. 
We refer to metallicity as overall metal content $Z$, i.e. the total mass of all the elements heavier than Helium 
that we track in our simulations (see Section~\ref{sec:CSF}) divided by the gas mass, and 
normalised to the Sun's metallicity. 
As for the Sun's metallicity, we adopt the present-day value $Z=0.01524$ \citep{Caffau2011}. 
Gas velocity in each bin is the mass-weighted velocity of gas particles in the bin: as a consequence, 
in star-forming regions where the bulk of gas is multiphase, the velocity estimate better reflects the 
velocity of warm and cold gas. 

The sequence of panels in Figure~\ref{IntroMapsZoom} shows that the central, quasar-host galaxy is 
embedded within a complex large-scale structure. It is located in the innermost region of a network 
of gaseous filaments which bridge surrounding galaxies and sub-structures, and shape 
the quasar-host galaxy environment. 
The central galaxy is fed by warm and cold gas which inflows from the large-scale environment. 
SF mainly occurs in the densest gas knots within the virial radius. 
The effect of past and ongoing SF is also visible in the distribution of heavy elements: 
gas metallicity ranges from $\sim 5 \cdot 10^{-3}$~Z$_{\odot}$ in rarefied gas far from the central galaxy 
to super-solar values of gas in and around sites of SF. 
Besides being responsible for metal enrichement of the ISM and circumgalactic medium, stellar feedback 
also promotes gas to outflow, along with quasar feedback. Radial velocities of the outflowing gas can even exceed 
$\sim 600$~km~s$^{-1}$ (see Figure~\ref{IntroMapsZoom} and Section~\ref{outflows_res}). 

Figure~\ref{IntroMaps4} introduces four among the simulations presented 
in Section~\ref{sec:simset}: {\sl BHs\_noFB}, {\sl AGN\_~fid}, {\sl SF\_only}, and {\sl SF\_only\_lowFB}. 
We show close-up views of the central galaxy in the four different simulations, 
as the focus of this work is the central galaxy hosted in the target sub-halo, and its ISM. 
For each simulated galaxy, we analyse gas density, temperature, metallicity, 
the SFR, and the mass-weighted, radial velocity of gas particles (as in Figure~\ref{IntroMapsZoom}). 
Maps of the galaxy model {\sl AGN\_~fid} in the second column represent a further zoom-in in the 
progressive view analysed in Figure~\ref{IntroMapsZoom}. 
We focus on the close-up views in Figure~\ref{IntroMaps4} to highlight differences between different models, 
the larger scale environment being almost indistinguishable among the considered runs 
(see also Appendix~\ref{OtherSims} and Figure~\ref{IntroMapsZoom_SFonly}). 
Before comparing results from Figures~\ref{IntroMapsZoom}~and~\ref{IntroMaps4} for different simulations, 
it is useful to introduce the SFH of the four systems.

\begin{figure}
\newcommand{\captionfonts}{\small}
\begin{minipage}{\linewidth}
\centering
\includegraphics[trim=0.1cm 0.3cm 0.2cm 0.0cm, clip, width=1.\textwidth]{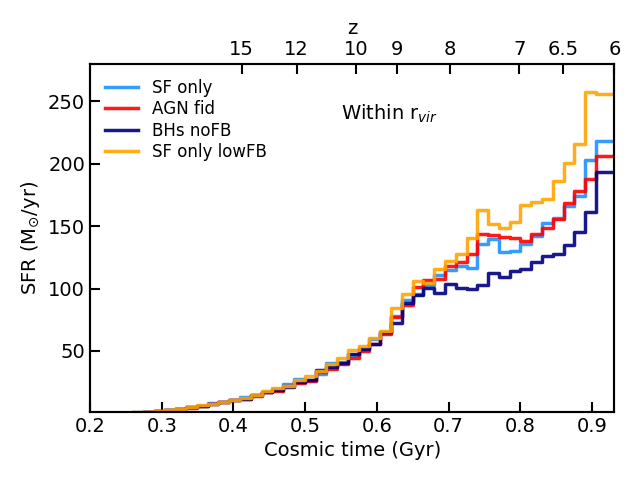}\\
\includegraphics[trim=0.1cm 0.1cm 0.2cm 0.2cm, clip, width=1.\textwidth]{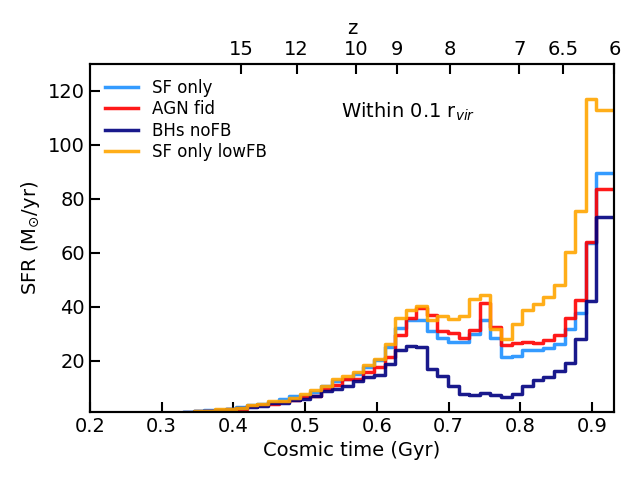}
\end{minipage} 
\caption{SFHs of the four galaxy models. 
	{\sl SF\_only}, {\sl AGN\_~fid}, 
	{\sl BHs\_noFB}, and {\sl SF\_only\_lowFB} 
	are shown in 
	light blue, red, blue, and orange, respectively.
	SFRs are computed by analysing star particles within 
	the virial radius {\sl (top panel)} and 
	one tenth of the virial radius {\sl (bottom panel)} of the most massive sub-halo in each simulation.}
\label{fig:sfr} 
\end{figure}

In Figure~\ref{fig:sfr} we show the SFH of the four simulated galaxies, within 
the virial radius and $0.1$~r$_{\rm vir}$ of the central, most massive halo at $z=6$. 
SFRs are retrieved by analysing stellar age distributions. 
SFRs of hundreds of~M$_{\odot}$~yr$^{-1}$ are in good agreement with observations 
of quasar-host galaxies at $z \sim 6$ \citep{Maiolino2005, Solomon2005, Wang2016, 
Trakhtenbrot2017a, Willott2017, Decarli2018, Venemans2018ApJ866}. 

All the simulations share a comparable SFH until $z \simeq 10$ is reached. 
Quasar feedback has only negligible effect on the SFH 
of the simulated galaxy in our model until $z=6$, 
and the simulations with ({\sl AGN\_~fid}) and without ({\sl SF\_only}) AGN feedback share comparable SFRs. 
The comparison between the two aforementioned models shows that the AGN can have both 
positive and negative feedback: when the quasar feedback is positive, the SFR can be enhanced because 
AGN feedback energy overpressurizes the ISM \citep[see][for details]{Valentini2020}. 
Episodes of negative quasar feedback (at $z \sim 6$) are due to the gas temperature increase induced by 
BH feedback on the surrounding gas.

\begin{table}
\centering
\begin{minipage}{85mm}
\caption[]{Virial radii and stellar masses for different simulations ({\sl {Column~1}}) at $z=6$. 
{\sl {Column~2:}} virial radius. 
{\sl {Column~3:}} stellar mass within~r$_{\rm vir}$. 
{\sl {Column~4:}} stellar mass within $0.1$~r$_{\rm vir}$.} 
\renewcommand\tabcolsep{4.mm}
\begin{tabular}{@{}lccc@{}}
\hline
Simulation  &  r$_{\rm vir}$   &  $M_{\ast} (r<$ r$_{\rm vir})$  &   $M_{\ast} (r< 0.1$ r$_{\rm vir})$     \\ 
                  &    (pkpc)           &    ($10^{10}$~M$_{\odot}$)     &  ($10^{10}$~M$_{\odot}$)   \\
\hline
\hline
{\sl AGN\_~fid}                &  $45.17$  &  $4.03$  & $1.13$     \\  
\hline
{\sl BHs\_noFB}              &  $44.95$  &   $3.51$ & $0.68$     \\  
\hline
{\sl SF\_only}                  &  $44.64$  &   $4.01$ & $1.07$     \\  
\hline
{\sl SF\_only\_lowFB}     &  $44.68$  &   $4.55$ & $1.44$     \\  
\hline
\hline
\end{tabular}
\label{tab:raggiEmasse}
\end{minipage}
\end{table}

\begin{figure*}
\newcommand{\captionfonts}{\small}
\begin{minipage}{\linewidth}
\centering
\includegraphics[trim=1.8cm 1.5cm 3.5cm 2.5cm, clip, width=.49\textwidth]{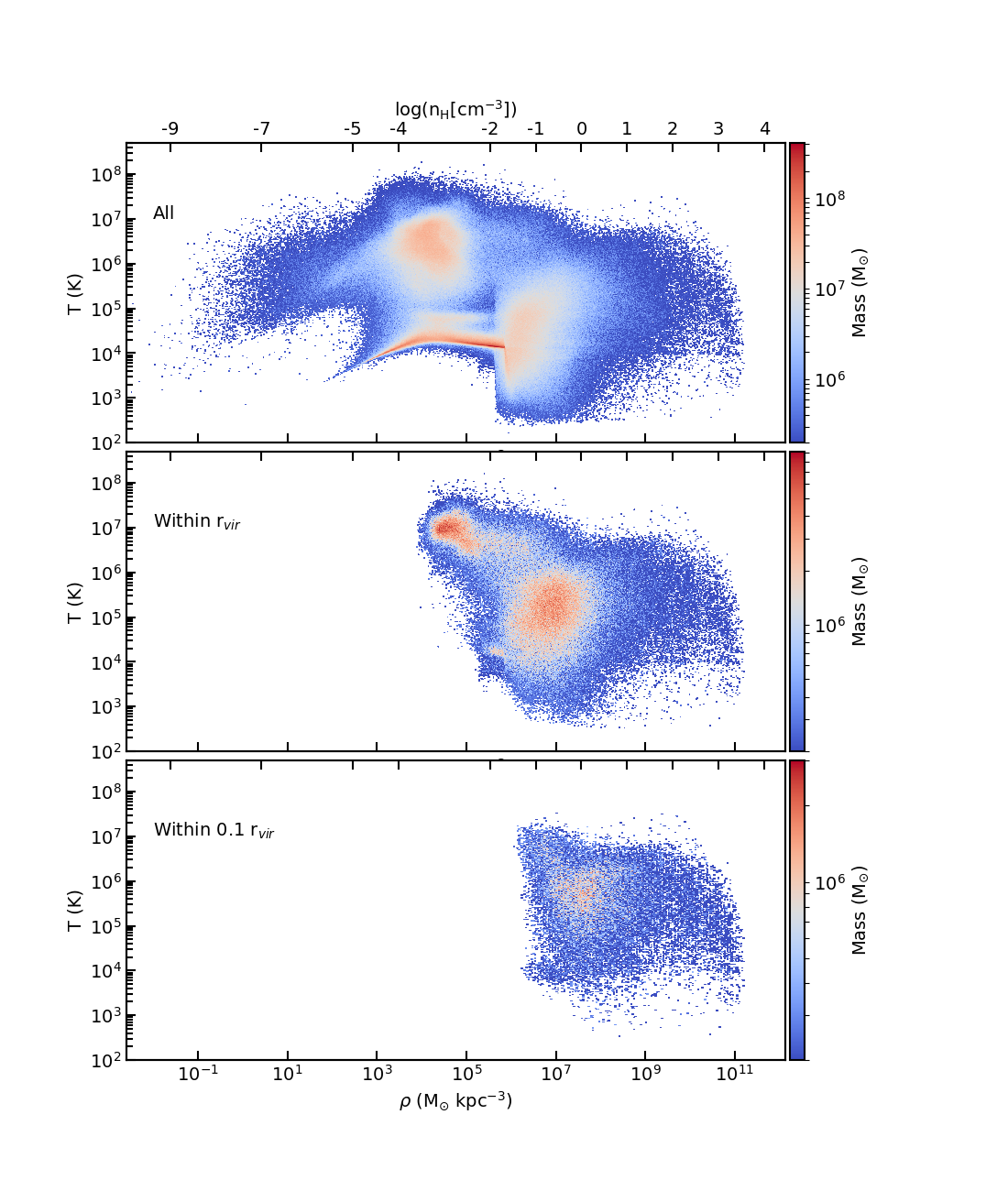} 
\includegraphics[trim=1.8cm 1.5cm 3.5cm 2.5cm, clip, width=.49\textwidth]{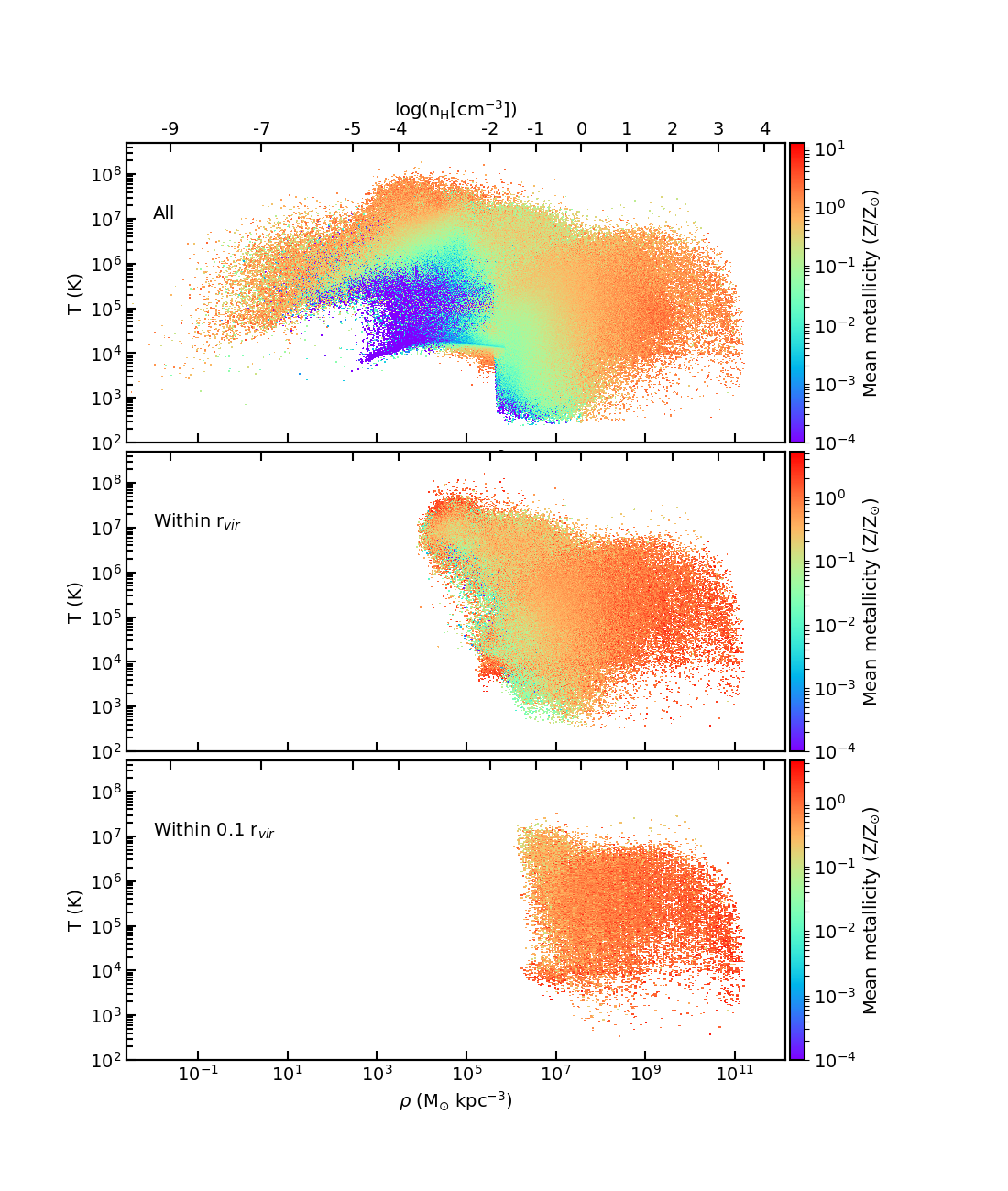}
\end{minipage} 
\caption{Distribution of gas particles in the density-temperature plane 
	in the reference simulation {\sl AGN\_~fid}, at redshift $z = 6$. 
	Top panels show the distribution of all the gas particles in the Lagrangian region,
	middle and bottom panels refer to gas particles within the virial radius r$_{\rm vir}$ and 
	within $0.1$~r$_{\rm vir}$, respectively. 
	The colour encodes the gas mass per density-temperature bin {\sl (left-hand panels)} and 
	the mean metallicity per bin {\sl (right-hand panels)}.
	All the color bars in the left set of panels share the minimum value, while the maximum of the color scale 
	is independent for each panel, to better capture features (the same is true for the three panels on the right).}
\label{PDs} 
\end{figure*}

The temperature of the ISM in the galaxy model {\sl AGN\_~fid} is on average higher 
than in {\sl SF\_only} (by a factor of $\sim 2$ for distances $r \lesssim 4$~kpc from the galaxy centre, 
as it can be seen by comparing the bottom panels of Figures~\ref{Profiles}~and~\ref{Profiles_SFonly}, 
and from Figure~\ref{IntroMaps4}). 
Moreover, the central region of the galaxy {\sl SF\_only} 
is more enriched in heavy metal 
($\gtrsim 3$~Z/Z$_{\odot}$, while the ISM has a solar metallicity in the innermost regions of 
the galaxy {\sl AGN\_~fid}, see Figure~\ref{IntroMaps4}), as a consequence of the higher SFR at $z=6$. 
Gas metallicities close to solar or super-solar in the innermost regions of high-redshift 
quasar-host galaxies are in agreement with observations 
\citep[e.g.][and references therein]{Juarez2009, Tang2019, Venemans2017_851}. 

As for the gas velocity in {\sl AGN\_~fid} and {\sl SF\_only}, Figure~\ref{IntroMaps4} shows that 
the quasar feedback is responsible for promoting outflowing gas to higher velocities 
(see also Section~\ref{outflows_res}). 
Also, the highest velocity gas outflowing from the innermost region of the quasar-host galaxy {\sl AGN\_~fid} 
has a more bipolar geometry with respect to gas outflowing in {\sl SF\_only}, 
the latter not showing a definite pattern. 
An enhanced bipolarity in the {\sl AGN\_~fid} model is due to the additional energy source 
represented by the central AGN. 
Interestingly, we find that the presence of the AGN is also responsible for disarranging the gas motion 
in the innermost regions of the galaxy 
(gas kinematics being less disturbed in {\sl SF\_only} than in {\sl AGN\_~fid}). 

When the central BH only accretes gas without injecting quasar feedback energy in the ISM ({\sl BHs\_noFB}), 
the SFR is lower with respect to all the other models. In this case, the central BH accretes more gas and 
grows way more massive with respect to the reference {\sl AGN\_~fid} model (see Section~\ref{BH_properties}): 
as a consequence, the central regions of the quasar-host galaxy lack fuel for SF as the gas 
is mainly accreted by the SMBH. 
The lower SFR in this model also stems from the higher temperature of the ISM (Figure~\ref{IntroMaps4}), 
and reflects on the lower gas metallicity. 

The model {\sl SF\_only\_lowFB} has the highest SFRs: when a lower kinetic stellar feedback efficiency 
is adopted (with respect to {\sl SF\_only}, see Section~\ref{sec:simset}), 
the velocity of particles receiving stellar feedback energy is boosted to lower values 
(see also Section~\ref{outflows_res}). Thus, a lower amount of gas is pushed far from sites of 
SF, and a larger reservoir of gas keeps fuelling SF in the galaxy. 
This galaxy model is also characterised by a higher gas metallicity, due to the higher SFR at $z \sim 6$. 

Virial radii of the central sub-halo in the four models, as well as 
the stellar mass enclosed within the virial radius and $0.1$~r$_{\rm vir}$ 
are listed in Table~\ref{tab:raggiEmasse}. 
Stellar masses of few $10^{10}$~M$_{\odot}$ in haloes as massive as $\sim 10^{12}$~M$_{\odot}$ 
at $z=6$ are also in line with predictions from stellar-to-halo mass relations obtained 
via abundance matching techniques \citep{Behroozi2013}.

\subsection{The host galaxy and its ISM}
\label{host_galaxies}

In this Section, we analyse the main features of the ISM of the quasar-host galaxy in our fiducial model.

Figure~\ref{PDs} shows the mass and metallicity distribution 
in the density-temperature phase diagram of gas particles in the simulation {\sl AGN\_~fid}, at redshift $z = 6$. 
The density of gas particles in Figure~\ref{PDs} is the SPH density; 
the temperature of gas particles is the SPH estimate for single-phase particles and 
the mass-weighted average of the temperatures of the hot and cold phases for multiphase particles. 
Multiphase particles in Figure~\ref{PDs} occupy the region where $\text{log}(n_{\rm H} [\text {cm}^{-3}]) > -2$ 
(corresponding to $n_{\rm H, \, thres}$; see Section~\ref{sec:CSF}). They scatter across an area that spans 
more than four orders of magnitude both in density and in temperature: 
such a spread is a characteristic feature 
of the advanced modelling of the ISM in our sub-resolution model. 
Indeed, the MUPPI sub-resolution model follows the dynamical evolution of the ISM 
and considers that the average energy of multiphase gas depends on its past history 
(the solution of the equations describing the ISM is not obtained under an equilibrium hypothesis). 
On the other hand, the spread in density and temperature 
would not be present if multiphase particles obeyed an equation of state \citep[e.g.][]{SpringelHernquist2003}, 
where the pressure $P$ of multiphase particles is a function of their density through 
a polytropic equation for $P(\rho)$ \citep[see discussion in][for details]{Valentini2017}. 

The right panels of Figure~\ref{PDs} show that the metallicity of gas spans more than five orders of magnitude, 
ranging from super-solar metallicity down to $\sim 10^{-4}$~Z$_{\odot}$. 
Extremely metal-poor gas is mainly warm ($\sim 5 \cdot 10^3 < T[K] < 10^6$) and 
rarefied ($-6 <\text{log}(n_{\rm H} [\text {cm}^{-3}]) < -2$). 
The ISM within the innermost region of the main galaxy (bottom right panel), on the other hand, 
has been significantly enriched by stellar evolution: its metallicity is rather homogeneous, 
and ranges from slightly sub-solar to super-solar.

\begin{figure}
\newcommand{\captionfonts}{\small}
\begin{minipage}{\linewidth}
\centering
\includegraphics[trim=0.5cm 0.cm 0.5cm 0.cm, clip, width=1.\textwidth]{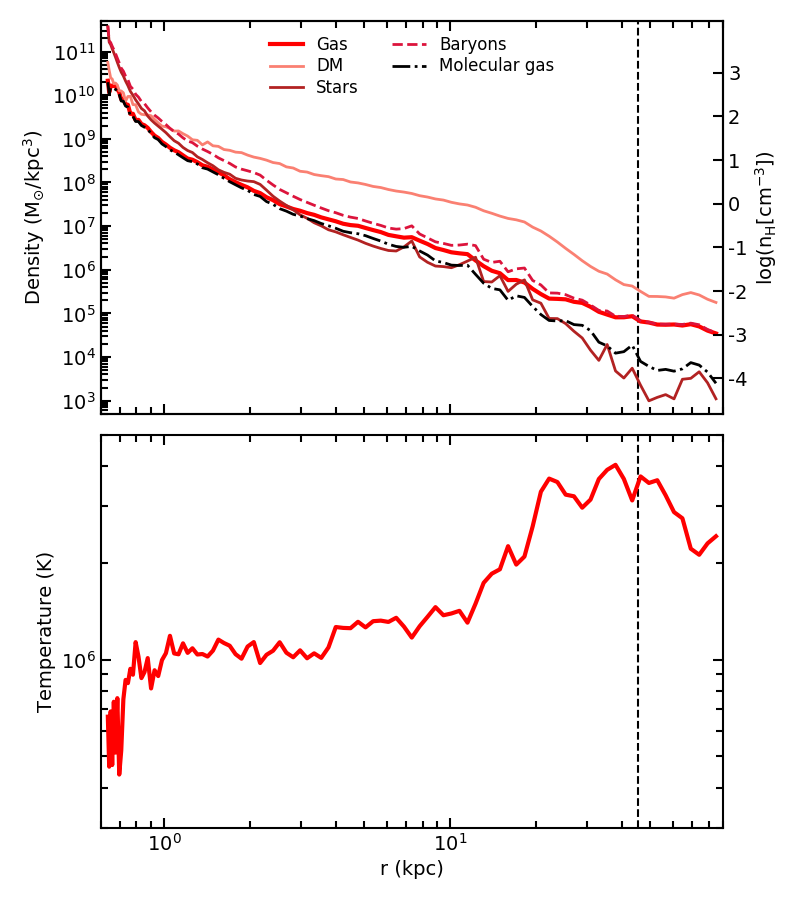} 
\end{minipage} 
\caption{Density and temperature radial profiles  
	within twice the virial radius 
	in the reference simulation {\sl AGN\_~fid}, at redshift $z = 6$.
	We show the density profile of gas (total and molecular), stars, DM, and baryons {\sl (top panel)}, 
	and the mass-weighted, temperature profile {\sl (bottom panel)}.
	The vertical, dashed black line highligths the virial radius of the most massive subhalo.}
\label{Profiles} 
\end{figure}

Figure~\ref{Profiles} shows radial profiles within twice the virial radius of the most massive subhalo 
in the reference simulation {\sl AGN\_~fid}, at redshift $z = 6$. 
In the top panel, we analyse density profiles of gas, stars, DM, and baryons, 
while the middle and bottom panels illustrate gas number density and 
mass-weighted temperature, respectively. 
This figure provides complementary information to Figure~\ref{PDs}, 
showing that the densest and coldest gas is located in the innermost regions of the quasar-host galaxy. 
The mean gas temperature increases from few~$10^5$~K in the centre 
to~$\sim 4 \cdot 10^6$~K at the virial radius, and then it mildly declines. 
The profile is not smooth, due to the presence of substructures and clumps, 
as it can be seen from Figure~\ref{IntroMapsZoom} and especially from Figure~\ref{IntroMaps} (right panel). 
These colder clumps are also responsible for the temperature decrease beyond the virial radius 
(see also Figure~\ref{IntroMapsZoom}, second row).

\begin{figure}
\newcommand{\captionfonts}{\small}
\begin{minipage}{\linewidth}
\centering
\includegraphics[trim=0.4cm 0.cm 0.2cm 0.cm, clip, width=1.\textwidth]{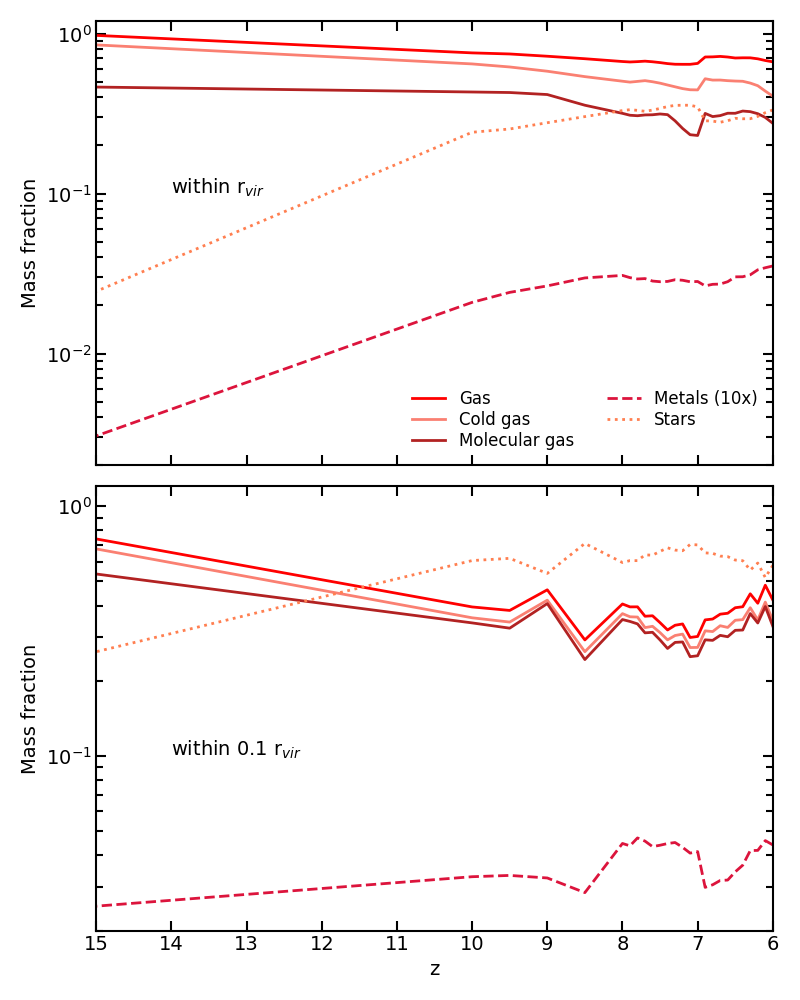} 
\end{minipage} 
\caption{Time evolution of gas, stellar, and metal mass 
	within the virial radius {\sl (top panel)} 
	and within $0.1$~r$_{\rm vir}$ {\sl (bottom panel)}
	in the reference simulation {\sl AGN\_~fid}, down to redshift $z = 6$. 
	We show the mass fraction (normalised to the total baryonic mass) 
	of all the gas, the cold gas ($T=300$~K), 
	the molecular gas, stars, and metals (multiplied by a factor of $10$).}
\label{Masses} 
\end{figure}

\begin{table}
\centering
\begin{minipage}{85mm}
\caption[]{Relevant masses of the quasar-host galaxy {\sl AGN\_~fid} 
within a given distance ({\sl {Column~1}}) from the centre. 
{\sl {Column~2:}} total mass of gas. 
{\sl {Column~3:}} mass of cold gas. Hot gas mass is $M_{\rm hot} = M_{\rm gas} - M_{\rm cold}$. 
{\sl {Column~4:}} mass of molecular gas. 
{\sl {Column~5:}} mass of metals in the gaseous phase. 
{\sl {Column~6:}} stellar mass.} 
\renewcommand\tabcolsep{1.05mm}
\begin{tabular}{@{}lccccc@{}}
\hline
Simulation  & $M_{\rm gas}$   &  $M_{\rm cold}$  &   $M_{\rm mol}$   & $M_{\rm metals}$    &    $M_{\ast}$         \\ 
{\sl AGN\_~fid}   &    ($10^{10}$~M$_{\odot}$)    &    ($10^{10}$~M$_{\odot}$)   &  ($10^{10}$~M$_{\odot}$)    &   ($10^8$~M$_{\odot}$)        &       ($10^{10}$~M$_{\odot}$)                                                   \\ 
at $z = 6$    &        &       &      &           &                                                        \\
\hline
\hline
within r$_{\rm vir}$ &  $8.05$  &   $4.91$ & $3.32$  &  $4.26$ &  $4.03$   \\  
\hline
within $0.1$~r$_{\rm vir}$ &  $0.82$  &  $0.68$  & $0.64$  &  $0.86$ &  $1.13$   \\  
\hline
\hline
\end{tabular}
\label{tab:masse}
\end{minipage}
\end{table}

\begin{figure*}
\newcommand{\captionfonts}{\small}
\begin{minipage}{\linewidth}
\centering
\includegraphics[trim=2.cm 0.cm 0.cm 0.7cm, clip, width=1.\textwidth]{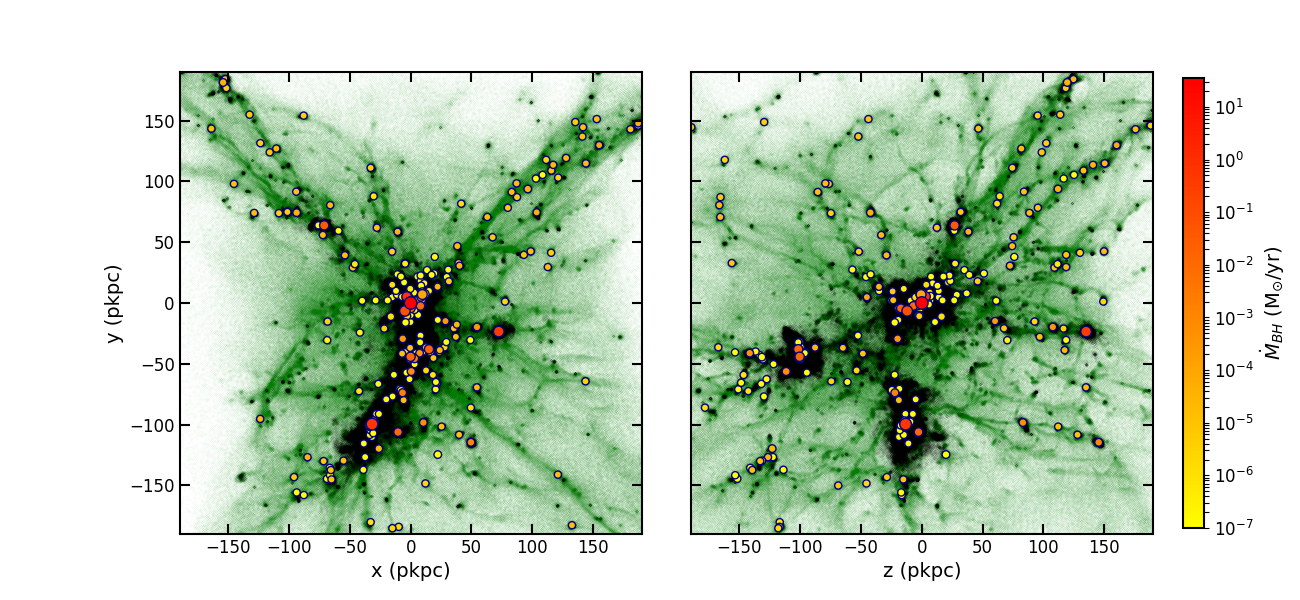} 
\end{minipage} 
\caption{Projected distribution of all the BHs in the reference simulation {\sl AGN\_~fid} at redshift $z = 6$, 
	in the planes $x-y$ {\sl (left-hand panel)} and $z-y$ {\sl (right-hand panel)}. 
	The colour bar encodes the BH accretion rate. 
	The size of each circle scales with the BH mass. 
	Distances are expressed in physical kpc (pkpc), 
	with respect to the position of the most massive BH, 
	which resides at the centre of the most massive subhalo. 
	Gas (green) and stellar (black) particle distributions are overlaid.}
\label{BHsMaps} 
\end{figure*}

Figure~\ref{Masses} shows the time evolution of the mass of gas, stars, and metals 
within the virial radius and $0.1$~r$_{\rm vir}$ 
of the most massive subhalo in the fiducial simulation {\sl AGN\_~fid}, at $z = 6$. 
In the figure we analyse the total amount of gas 
(contributed by all the gas particles within either r$_{\rm vir}$ or $0.1$~r$_{\rm vir}$), 
the mass of cold gas (which constitutes the bulk of the mass of multiphase gas particles; see Section~\ref{sec:CSF}), 
and the molecular gas mass (representing a fraction of the mass of cold gas, as detailed in Section~\ref{sec:CSF}). 
We also consider the mass of metals (contributed by all the heavy elements within gas particles), 
and the stellar mass. 
The mass of gas (and thus that of metals and stars) smoothly increases across the (almost) entire time frame, 
as a consequence of the gas which is accreted from the large scale structure and which provides the reservoir 
for SF. 
Focussing on the evolution within $0.1$~r$_{\rm vir}$ (bottom panel) at $9 > z > 8.5$, 
it is possible to see that the amount of gas decreases, as a consequence of the gas expelled beyond $0.1$~r$_{\rm vir}$ 
by galactic outflows triggered by ongoing SF (see Figure~\ref{fig:sfr}). 

Figure~\ref{Masses} also illustrates how significant the contribution of the cold and molecular phases is 
to the total amount of gas, especially within $0.1$~r$_{\rm vir}$. 
When the ISM is almost entirely multiphase, the cold gas accounts for $\sim 85 \%$ of the total gas, 
while the hot phase contributes little. 
As for the amount of cold gas which is in the molecular phase, it depends on the ISM properties 
(i.e. gas pressure and thus density) through the molecular fraction $f_{\rm mol}$ (see Section~\ref{sec:CSF}). 
While this fraction spans the entire range of values when considering multiphase gas within the virial radius, 
it easily approaches $\sim 0.8 - 1$ (for the majority of gas particles) within $0.1$~r$_{\rm vir}$, 
where the ISM is denser and more pressurised due to the activity of AGN and especially stellar feedback. 
As a consequence, the mass of molecular gas is close to that of the cold gas. 
Masses of gas (total, cold, and molecular), metals and stars 
within the virial radius and $0.1$~r$_{\rm vir}$ of the quasar-host galaxy {\sl AGN\_~fid} at $z = 6$ 
are listed in Table~\ref{tab:masse}.

\begin{table}
\centering
\begin{minipage}{85mm}
\caption[]{Most massive BH and subhalo properties for different simulations ({\sl {Column~1}}) at $z=6$. 
{\sl {Column~2:}} most massive, central BH mass. 
{\sl {Columns~3~and~4:}} most massive BH accretion rate, in units of M$_{\odot}$/yr and 
	Eddington accretion rate, respectively. 
{\sl {Column~5:}} stellar mass of the subhalo which hosts the most massive BH 
	(as identified by the SUBFIND algorithm).} 
\renewcommand\tabcolsep{2.9mm}
\begin{tabular}{@{}lcccc@{}}
\hline
Simulation & $M_{\rm BH}$     & \multicolumn{2}{c}{$\dot{M}_{\rm BH}$}    & $M_{\ast}$ \\ 
                 & (M$_{\odot}$)      &  (M$_{\odot}$ / yr) & ($\dot{M}_{\rm Edd}$)  & ($10^{10}$~M$_{\odot}$) \\
\hline
\hline
{\sl AGN\_~fid}                &  $9.85 \cdot 10^8$  &  $35.53$  &    $0.495$   & $2.63$     \\  
\hline
{\sl BHs\_noFB}              &  $4.62 \cdot 10^{11}$  & $3.17 \cdot 10^4$ &    $0.978$   &    $2.26$     \\  
\hline
{\sl AGN\_highFB}          &  $4.16 \cdot 10^7$  &  $8.53 \cdot 10^{-2}$   &     $0.028$   &  $2.64$     \\  
\hline
\hline
\end{tabular}
\label{tab:BHs_subhaloMass}
\end{minipage}
\end{table}

\begin{table}
\centering
\begin{minipage}{85mm}
\caption[]{Main features of the 2 most accreting BHs after the central, most massive one (columns~2~and~3)
and of the two closest BHs to the most massive one (columns~4~and~5) in the simulation {\sl AGN\_~fid} at $z=6$. 
{\sl {Row~1:}} mass. 
{\sl {Row~2:}} accretion rate. 
{\sl {Row~3:}} distance from the most massive BH. } 
\renewcommand\tabcolsep{2.3mm}
\begin{tabular}{@{}lcccc@{}}
\hline
Simulation                                   &   \multicolumn{2}{c}{Most accreting BHs}      & \multicolumn{2}{c}{Closest BHs}     \\ 
{\sl AGN\_~fid}  $\,$  ($z = 6$)                        &  2nd    &   3rd                           &    1st    &    2nd   		       \\ 
\hline
\hline
$M_{\rm BH}$ (M$_{\odot}$)          & $9.34 \cdot 10^6$  &  $2.71 \cdot 10^8$        &  $1.48 \cdot 10^5$  &   $1.48 \cdot 10^5$      \\  
\hline
$\dot{M}_{\rm BH}$ (M$_{\odot}$ / yr)              & $0.66$  &  $0.55$                        &  $5.5 \cdot 10^{-7}$  &  $2.7 \cdot 10^{-7}$  	 \\  
\hline
d (pkpc) 			   				  & $6.87$  &  $105.4$                       &  $0.023$  &  $0.024$  				    \\  
\hline
\hline
\end{tabular}
\label{tab:closestBHs}
\end{minipage}
\end{table}

\subsection{BH properties}
\label{BH_properties}

In this Section, we discuss the properties of BHs in our reference simulation. 
Figure~\ref{BHsMaps} introduces the distribution of all the BHs in the reference simulation {\sl AGN\_~fid} 
at $z = 6$, along with gas and stars. 
BHs are color-coded according to their accretion rate, while  
the size of each circle scales with the BH mass. 
BH masses range from the adopted seed value 
($M_{\rm BH, \, seed}=1.48 \cdot 10^5$~M$_{\odot}$, see Section~\ref{sec:AGN}) 
to the most massive BH formed $M_{\rm BH} = 9.85 \cdot 10^8$~M$_{\odot}$. 
There are $2$ BHs more massive than $10^8$~M$_{\odot}$ in the simulation {\sl AGN\_~fid}, 
$3$ BHs whose mass is in the range $10^7 - 10^8$~M$_{\odot}$, 
and $9$ BHs with mass between $10^6$ and $10^7$~M$_{\odot}$. 
The accretion rate of the most massive BH is $\dot{M}_{\rm BH} = 35.53$~M$_{\odot}$/yr 
(see Table~\ref{tab:BHs_subhaloMass}). 
The most massive BH was seeded at $z=12.53$ and has since then experienced $30$ mergers with other BHs.
The last $8$ mergers experienced occurred between $6.1 < z < 6$.
The main properties of the two most accreting BHs after the central, most massive one 
and those of the two closest BHs to the most massive one at $z=6$
are listed in Table~\ref{tab:closestBHs}. 
The two closest BHs are BHs which have just been seeded.

\begin{figure}
\newcommand{\captionfonts}{\small}
\begin{minipage}{\linewidth}
\centering
\includegraphics[trim=0.2cm 0.cm 0.2cm 0.cm, clip, width=1.\textwidth]{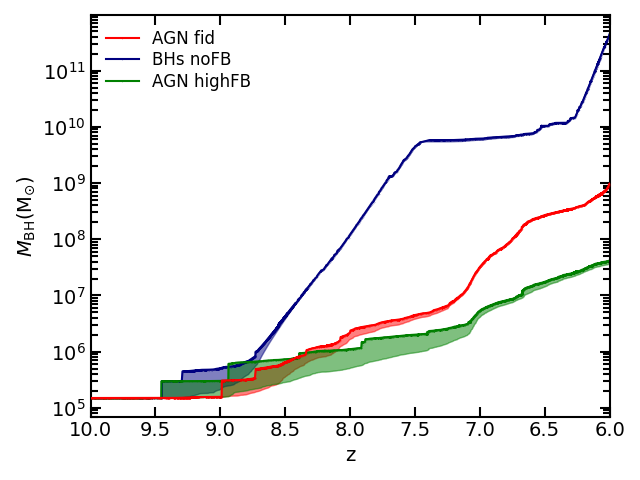} 
\end{minipage} 
\caption{Time evolution of the BH mass for the central, most massive BH in the simulations
{\sl AGN\_~fid} (red curve), 
{\sl BHs\_noFB} (blue), and
{\sl AGN\_highFB} (green). 
The shaded region for each curve shows the contribution to BH mass growth from mergers with other BHs, 
which is negligible with respect to gas accretion, at $z=6$. 
The final mass of the BH in {\sl AGN\_~fid} is in agreement with observations at $z=6$ (see Figure~\ref{Magorrian}), 
while the final mass of the model {\sl AGN\_highFB} is in line with the $M_{\rm BH} - M_{\star}$ relation 
observed at low redshift.}
\label{BHmassGrowth} 
\end{figure}

\subsubsection{SMBH mass growth}
\label{BHmasses}

Figure~\ref{BHmassGrowth} shows the BH mass growth as a function of the redshift 
for the central, most massive BH in three simulations of our suite. 
Besides {\sl AGN\_~fid} and {\sl BHs\_noFB}, here we also consider results from the simulation {\sl AGN\_highFB}. 
This simulation (see Section~\ref{sec:simset} and Table~\ref{tab:otherSims}) 
is analogous to the fiducial {\sl AGN\_~fid} but adopts a feedback efficiency higher\footnote{
We further note that the simulation {\sl AGN\_highFB} adopts a feedback efficiency ($\epsilon_{\rm f} = 0.001$) that is ten times lower than the reference model of \citealt[][]{Valentini2020} (i.e. $\epsilon_{\rm f} = 0.01$). This is because, while in the latter simulation BHs have more than $13$~Gyr to grow supermassive, in the former they have less than $1$~Gyr to reach masses that are in agreement with observations in the local universe. As a consequence, a weaker AGN feedback is required.} 
than the reference value by a factor of $10$. 

Figure~\ref{BHmassGrowth} also quantifies the contributions to BH mass growth from the two possible channels, 
namely gas accretion and mergers with other BHs. The shaded region underlying each of the three curves shows 
the BH mass increase due to BH mergers. 
Thus, the lower border of the shaded area shows the mass that the BH would have 
if it only grew because of gas accretion. 
By analysing the shaded area we can thus appreciate the marginal contribution of BH-BH merger 
to the increase of BH mass. This negligible contribution mainly stems from the fact that the central SMBH experiences 
mergers with BHs whose mass is by far smaller than its own.

Table~\ref{tab:BHs_subhaloMass} lists mass and accretion rate of the central BH in {\sl BHs\_noFB}. 
Although the AGN feedback in our model does not impact significantly on the physical properties of the ISM 
of the quasar-host galaxy, it has a key role in regulating the SMBH mass growth. 
Albeit a tiny amount ($\epsilon_{\rm r} \cdot \epsilon_{\rm f} = 3 \cdot 10^{-6}$) of the rest-mass energy 
that the BH accretes is coupled to the ISM, this AGN feedback energy is crucial to avoid that the SMBH 
grows way too massive (the inclusion of AGN feedback reduces the final BH mass by a factor of $\sim 400$). 

\begin{figure}
\newcommand{\captionfonts}{\small}
\begin{minipage}{\linewidth}
\centering
\includegraphics[trim=0.55cm 0.cm 0.3cm 0.cm, clip, width=1.04\textwidth]{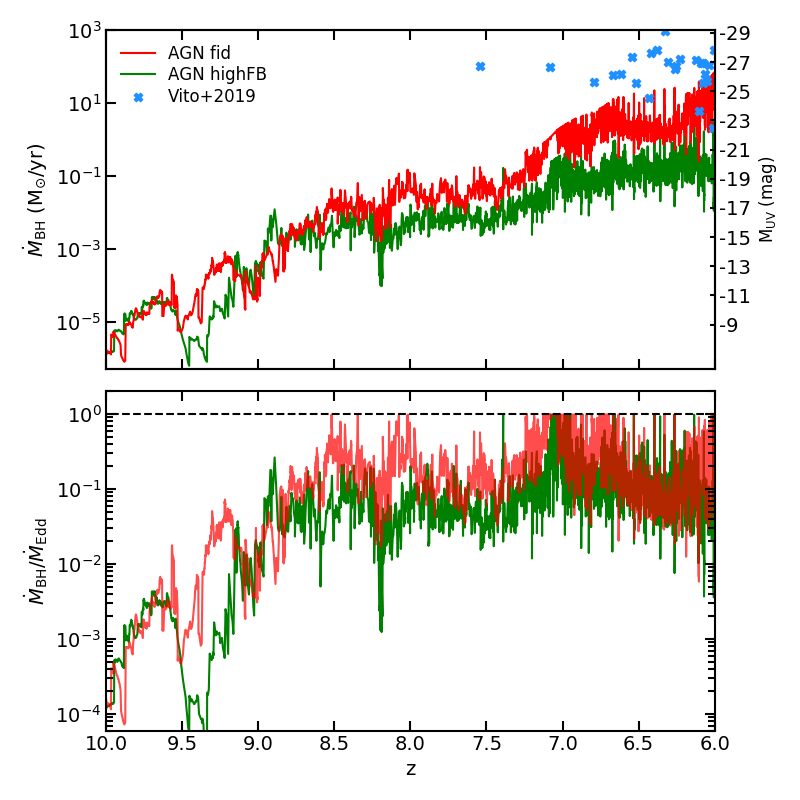} 
\end{minipage} 
\caption{Evolution of the accretion rate of the most massive BH 
	in the simulations {\sl AGN\_~fid} and {\sl AGN\_highFB}.  
	The same evolution is shown both in units of~M$_{\odot}$/yr (top panel) and
	in units of the Eddington accretion rate (bottom panel). The dashed
	black line where $\dot{M}_{\rm BH} / \dot{M}_{\rm Edd} = 1$ marks the 
	maximum allowed BH accretion rate in our models.
	Observational data from \citet{Vito2019}.}
\label{BHmdot} 
\end{figure}

Main features of the central, most massive BH in {\sl AGN\_highFB} 
and its host subhalo are listed in Table~\ref{tab:BHs_subhaloMass}, and 
are in agreement with the $M_{\rm BH} - M_{\star}$ relation inferred from low-redshift observations (see below). 
By adopting the following relation\footnote{We adapted the relation from \citet{DiMascia2021} 
	by considering the value of the radiative efficiency adopted in our simulations 
	($\epsilon_{\rm r}=0.03$ instead of the commonly assumed $0.1$).} 
$M_{\rm UV} = -21.7 -2.5 \text{ log}_{\rm 10} \dot{M}_{\rm BH}$
to estimate the intrinsic, dust unabsorbed UV magnitude of the quasar 
as a function of the BH accretion rate (in units of~M$_{\odot}$/yr),
we obtain 
$M_{\rm UV} = - 25.6$ for the most massive BH in {\sl AGN\_~fid} at $z=6$, 
while $M_{\rm UV} = -19$ in {\sl AGN\_highFB} 
(see Figure~\ref{BHmdot}, also to appreciate how variable BH accretion rates 
and hence AGN luminosities are).  
This low luminosity for the SMBH in {\sl AGN\_highFB} 
\citep[that would not be in agreement with the luminosity of the quasar sample by][]{Matsuoka2018} 
supports the need to calibrate BH physics 
according to high-redshift observations (see Sections~\ref{BHaccRate}~and~\ref{BHmago})
to study high-redshift quasars in simulations.

As for the role played by the adopted angular momentum dependent gas accretion onto SMBHs 
in determining final BH masses, we have thoroughly investigated this process in \citet{Valentini2020}.  
We find that when the accretion of cold gas that is supported by rotational velocity is diminished 
via equation~(\ref{Mdot_limited_mango}), the evolution of the BH mass changes: 
reducing the accretion of cold gas delays and decreases the BH growth. 
The higher the values of $C_{\rm visc}$ that are adopted (equation~(\ref{AngMomLimiter})), 
the more significant is the BH growth reduction. 
The impact of different values of $C_{\rm visc}$ is thoroughly quantified in \citet{Valentini2020} 
(see in particular Section 5.6, and Figures~16~and~17 of that paper).

\begin{figure*}
\newcommand{\captionfonts}{\small}
\begin{minipage}{\linewidth}
\centering
\includegraphics[trim=0.8cm 0.cm 0.4cm 0.cm, clip, width=.49\textwidth]{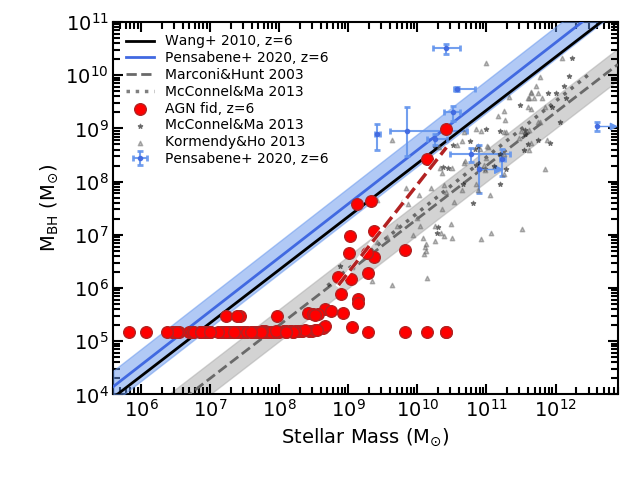}
\includegraphics[trim=0.8cm 0.cm 0.4cm 0.cm, clip, width=.49\textwidth]{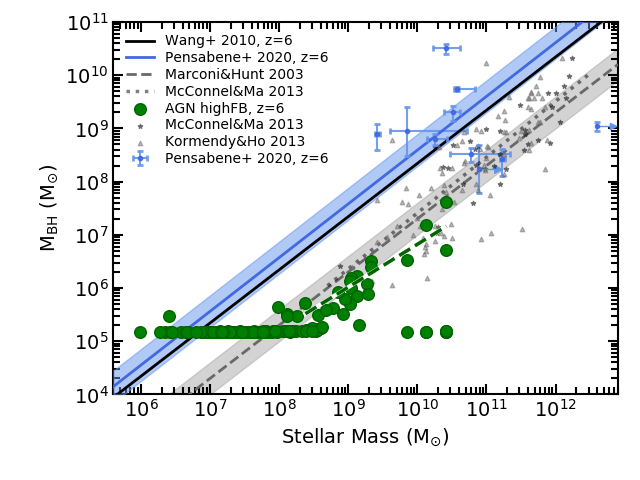}
\end{minipage} 
\caption{$M_{\rm BH} - M_{\star}$ relation for the reference simulation {\sl AGN\_~fid} ({\sl left}) 
	and for the model {\sl AGN\_highFB} ({\sl right}), at redshift $z = 6$. 
	Red and green circles are results from the two simulations, 
	the dashed line (same colour) showing the best fit for each model. 
	We also show best-fit relations from high-redshift observations by \citet{Wang2010} (solid black line) 
	and by \citet{Pensabene2020} (light-blue line, with the shaded region highlighting the 1$\sigma$ uncertainty 
	and the light-blue symbols representing the sample used to infer the best fit). 
	Dashed and dotted grey lines show the local $M_{\rm BH} - M_{\star}$ relations 
	inferred by \citet{Marconi2003} (the shaded envelope being the scatter around it) 
	and by \citet{McConnell2013} (using a sample of early-type galaxies, pinpointed by grey stars). 
	Grey triangles are observations by \citet{KormendyHo2013} for low-redshift ellipticals and late-type galaxies.}
\label{Magorrian} 
\end{figure*}

\subsubsection{SMBH accretion rates}
\label{BHaccRate}

The evolutions of the most massive BH accretion rate in {\sl AGN\_~fid} and {\sl AGN\_highFB} 
are presented in Figure~\ref{BHmdot}.  
The top panel describes the redshift evolution of the accretion rate in units of~M$_{\odot}$/yr, 
while the bottom panel shows the same evolution in units of the Eddington accretion rate. 
The BH accretion rate is capped to the Eddington accretion rate (dashed line in Figure~\ref{BHmdot}; see Section~\ref{sec:AGN}) in our simulations.  
We also include observational data from the high-z quasar sample of \citet{Vito2019}: 
we converted observed UV magnitudes in BH accretion rates 
by exploiting the same relation used in Section~\ref{BHmasses}. 
The top panel of Figure~\ref{BHmdot} shows how the accretion rate increases as the redshift decreases; 
by focussing on the bottom panel, it is possible to see that the two BHs experience 
an early phase ($z \gtrsim 9$) of low-accretion rate and then they enter a higher-accretion rate stage.  
The commonly adopted threshold to distinguish between high- and low-accretion rate mode feedback 
is $\dot{M}_{\rm BH} / \dot{M}_{\rm Edd} = 10^{-2}$ \citep[e.g.][]{Churazov2005, Sijacki2007}. 
For redshifts lower than $z\sim 9$, the most massive BH in {\sl AGN\_~fid} always accretes at 
high-accretion rates (quasar mode). 
In particular, the SMBH is characterised by several episodes where its accretion is Eddington limited. 
Throughout the BH evolution, the accretion of cold gas always dominates over the accretion of the hot gas, 
and it almost amounts to the total BH accretion rate (see equation~(\ref{Mdot_limited})).
At $z=6$, the accretion rate (in units of the Eddington accretion rate) 
of the most massive BH in {\sl AGN\_~fid} is $\dot{M}_{\rm BH} / \dot{M}_{\rm Edd} = 0.495$. 
This value is significantly lower for the central BH 
in {\sl AGN\_highFB}, where $\dot{M}_{\rm BH} / \dot{M}_{\rm Edd} = 2.77 \cdot 10^{-2}$ 
highlights an AGN activity which is at the limit of the quasar phase (according to the aforementioned criterion). 
There are way fewer episodes of Eddington-limited accretion in {\sl AGN\_highFB} 
than in {\sl AGN\_~fid}.
Hence, we find that SMBHs on the local $M_{\rm BH} - M_{\star}$ relation 
have accretion rates which are lower than those characterising high-z quasars 
\citep[e.g.][see also Table~\ref{tab:otherSims}]{Mazzucchelli2017, Vito2019}.

\subsubsection{The $M_{\rm BH} - M_{\star}$ relation}
\label{BHmago}

Figure~\ref{Magorrian} shows the $M_{\rm BH} - M_{\star}$ relation. 
We analyse results for the reference simulation {\sl AGN\_~fid} (red points, on the left) 
and for the model {\sl AGN\_highFB} (green, on the right), at redshift $z = 6$. 
Each cirle pinpoints a BH in the simulation as a function of the stellar mass of the subhalo in which the BH resides 
(as provided by the SUBFIND algorithm). We consider only those BHs whose distance from the centre 
of their host subhalo is smaller than twice the half-mass radius of the subhalo itself (not to include BHs wandering because of spurious, numerical effects; see Section~\ref{sec:AGN}). 
Assuming a linear relation of $\text{log}(M_{\rm BH})$ with $\text{log}(M_{\ast})$, 
we find the following best-fit parameters (considering BHs with $M_{\rm BH} > 5 \cdot 10^5$~M$_{\odot}$): 
$\text{log}(M_{\rm BH}) = 1.656 \, \text{log}(M_{\ast}) - 8.615$ for {\sl AGN\_~fid}, and 
$\text{log}(M_{\rm BH}) = 0.806 \, \text{log}(M_{\ast}) - 1.237$ for {\sl AGN\_highFB}. 

The mass of the central, most massive BH in our models 
and the stellar mass of its host subhalo are listed in Table~\ref{tab:BHs_subhaloMass}. 
The lowest mass BHs in Figure~\ref{Magorrian} are BHs whose mass corresponds 
to the seed value (see Section~\ref{sec:AGN}). 
As discussed in Sections~\ref{sec:AGN}~and~\ref{sec:simset}, 
we calibrated the feedback efficiency of the AGN model in our fiducial run so that the final mass 
of the most massive BH was large enough to meet the $M_{\rm BH} - M_{\star}$ relation observed at high-redshift. 

We compare predictions from our simulations to observations at $z = 6$ and in the local universe. 
As for high-redshift observations, we show best-fit relations found by \citet{Wang2010} 
and by \citet{Pensabene2020}. 
The normalization of the $M_{\rm BH} - M_{\star}$ relation in the low-redshift universe is lower than that inferred 
from observations at $z=6$ by a factor of~$\sim15$ \citep[][see also Section~\ref{sec:discussion}]{Wang2010}. 

At high redshift, the $M_{\rm BH} - M_{\star}$ relation in our simulations is shaped entirely by quasar feedback, 
which controls the BH growth while leaving SF almost unaffected 
(due to its inability to hamper the cosmological infall, see Section~\ref{outflows_res}). 
The slope of the $M_{\rm BH} - M_{\star}$ relation inferred from our reference simulation {\sl AGN\_~fid} 
is steeper than that suggested by observations in the local universe \citep[and usually assumed when inferring 
the normalization of this relation with high-z data, e.g.][]{Pensabene2020}. 
This implies that lower mass BHs experience a mass growth which is not as fast as suggested by observations 
at low redshift.  
We note that BHs in the model {\sl AGN\_highFB} at $z > 6$ do not shape a $M_{\rm BH} - M_{\star}$ relation 
with a slope considerably steeper than that at $z=6$, 
nor comparable with that characterising the model {\sl AGN\_~fid} at $z=6$. 
This suggests that it is unlikely that BHs lying on a $M_{\rm BH} - M_{\star}$ relation whose slope is in agreement 
with that suggested by observations in the local universe have undergone a stage in which BHs of different mass 
were growing at a different pace. 

Physical processes occurring on scales relevant for the BH accretion process may be responsible for 
the aforementioned trend we find in the {\sl AGN\_~fid} model. 
On the other hand, processes not included in our simulations may represent a caveat for our findings.
For instance, if the quasar radiative efficiency $\epsilon_{\rm r}$ depended on the BH spin 
and thus increased with the BH mass \citep[e.g.][]{DavisLaor2011}, 
lower mass BHs would have a smaller efficiency and grow more because of less quasar feedback. 
Upcoming data are needed and crucial to confirm the possible deviation from the slope 
of the local $M_{\rm BH} - M_{\star}$ relation suggested by our fiducial model.

\subsection{Inflow and outflow}
\label{outflows_res}

In this Section, we analyse properties of inflowing and outflowing gas in four simulations 
introduced in Section~\ref{sims_overview}.

\begin{figure*}
\newcommand{\captionfonts}{\small}
\begin{minipage}{\linewidth}
\centering
\includegraphics[trim=0.1cm 0.cm 0.1cm 0.cm, clip, width=1.\textwidth]{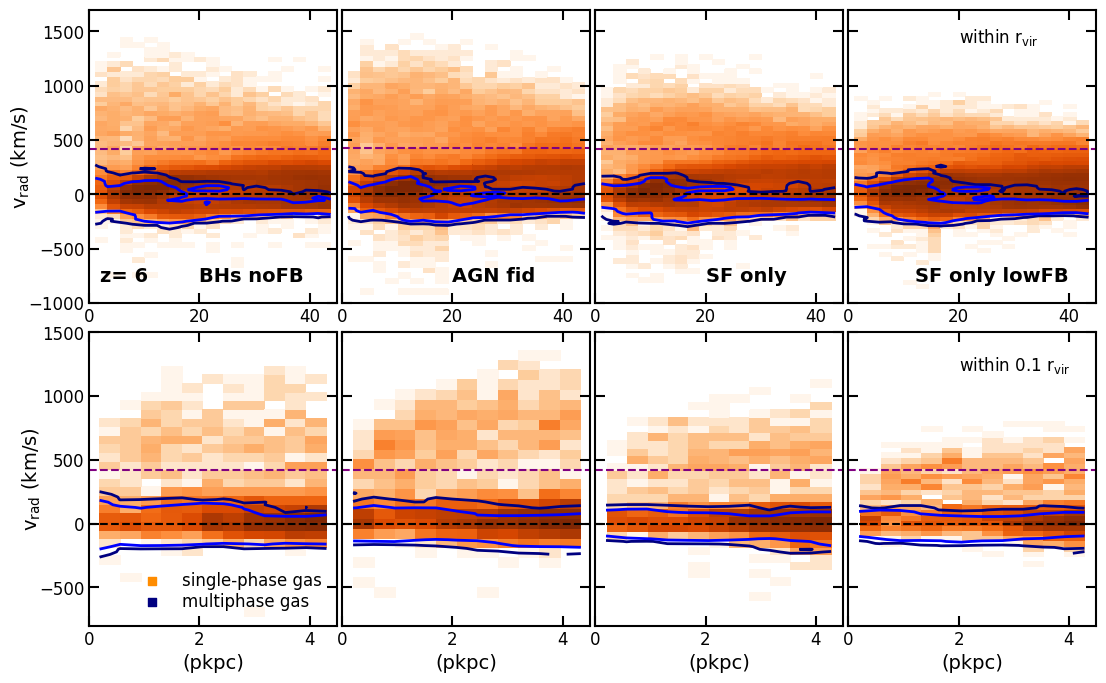} 
\end{minipage} 
\caption{Radial velocity for single-phase gas particles (i.e. hot gas) and 
	multiphase particles (i.e. cold gas, see the text for details) 
	as a function of the distance from the centre of the main halo, at $z=6$. 
	The background histogram shows the distribution of single-phase gas, the colour encoding 
	the fraction of particles in each bin with respect to the total number of single-phase particles 
	(darker shades pinpoint bins with a larger number of particles). 
	Blue contours overlapping the background histogram show the distribution of multi-phase gas.  
	From left to right we consider models: {\sl BHs\_noFB}, {\sl AGN\_~fid}, 
	{\sl SF\_only}, and {\sl SF\_only\_lowFB}. 
	We consider gas within the virial radius ({\sl top panels}), 
	and within $0.1$~r$_{\rm vir}$ ({\sl bottom panels}). 
	The horizontal, dashed black line marks $v_{\rm rad}= 0$~km/s, thus separating inflow from outflow. 
	For each model, the horizontal, dashed purple line highlights the escape velocity of the halo. 
	Escape velocities range from~$418.4$~to~$423.4$~km/s, according to the model. }
\label{Outflows} 
\end{figure*}

Figure~\ref{Outflows} shows the distribution of radial velocities of gas as a function of the distance from the 
centre of the main subhalo for the different models, at $z=6$. 
We consider gas within the virial radius (upper panels) 
and within $0.1$~r$_{\rm vir}$ (lower panels). 
We show that gas which is outflowing (i.e. which has a positive radial velocity $v_{\rm rad}$) 
can reach velocities as high as $\sim 1500$~km/s should the quasar feedback be included 
(model {\sl AGN\_~fid}), while velocities are on average lower 
when only the stellar feedback is accounted for (models {\sl SF\_only} and {\sl SF\_only\_lowFB} -- see also below). 
As for the simulation {\sl BHs\_noFB}, the enhanced (compared to e.g.  {\sl SF\_only})
velocities of outflowing gas are due 
to the central SMBH which increases the gravitational potential and heats the gas up to higher temperatures 
(see Figure~\ref{IntroMaps4}). In addition, the ISM in this latter model has not been enriched in heavy elements as 
in other models (due to a lower SFR, see Figure~\ref{fig:sfr}), and thus it is easier for stellar feedback 
to accelerate it up to larger speeds. 

We distinguish between single-phase and multiphase outflowing/inflowing gas. 
Single-phase gas within~r$_{\rm vir}$ has a temperature ranging between $10^6 - 10^8$~K 
(see for instance Figure~\ref{PDs}, middle left panel), while the bulk of multiphase gas 
has a temperature $\leq 10^5$~K; in particular, cold ($T_{\rm c}=300$~K, see Section~\ref{sec:CSF}) and molecular gas 
represent almost $\sim90 \%$ of the mass budget of multiphase particles in our model (see Figure~\ref{Masses}), 
so it is possible to identify cold gas in Figure~\ref{Outflows} with gas whose temperature is of few hundreds~K.
The figure illustrates that different phases have different kinematics: the hot and diffuse gas has 
higher velocities, which can easily exceed the escape velocity of the halo; multiphase gas is characterised by 
lower velocities, only in a few cases exceeding $\sim 300$~km/s, 
and makes up for the almost totality of the inflowing gas. 
Escape velocities for the different models are listed in Table~\ref{tab:GasOutflow}. 
We stress that the velocity of gas in our model, 
in good agreement with observations (see below), 
is the result of the modelling of feedback processes and of the advanced treatment of SPH included 
in our simulations. Indeed, within our feedback prescriptions, we do not assume any ad hoc wind velocity, nor 
we kick particles to a defined velocity suggested by observations or theoretical models 
\citep[see][for further details]{muppi2014, Valentini2017, Valentini2020}.

\begin{figure*}
\newcommand{\captionfonts}{\small}
\begin{minipage}{\linewidth}
\centering
\includegraphics[trim=0.1cm 0.cm 0.1cm 0.cm, clip, width=1.\textwidth]{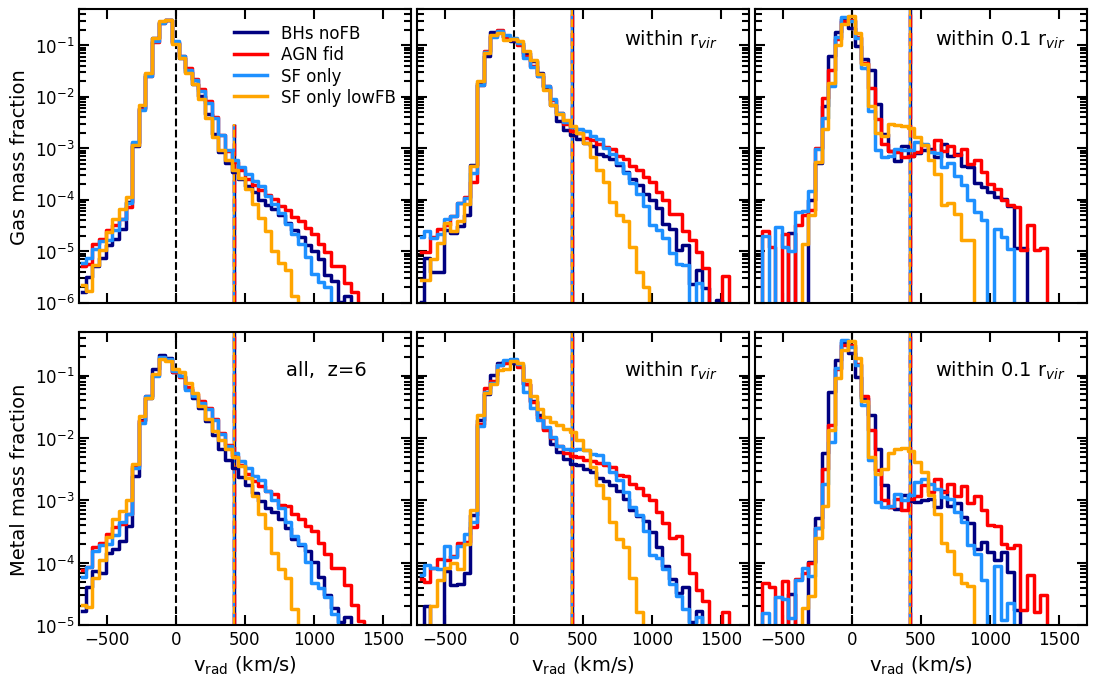} 
\end{minipage} 
\caption{Histogram of the radial velocity of gas (in terms of gas mass fraction -- {\sl top panels}, 
	and metal mass fraction -- {\sl bottom panels})
	within different regions around the main halo, at $z=6$. 
	We consider the four models {\sl BHs\_noFB} (blue), {\sl AGN\_~fid} (red), 
	{\sl SF\_only} (light blue), and {\sl SF\_only\_lowFB} (orange). 
	We analyse: 
	all the gas particles in the volume ({\sl left panels}), 
	gas within the virial radius ({\sl middle panels}), and 
	gas within $0.1$~r$_{\rm vir}$ ({\sl right panels}).  
	The vertical, dashed black line marks $v_{\rm rad}= 0$~km/s, hence distinguishing between inflow and outflow. 
	For each model, the vertical line highlights the escape velocity of the halo (colours as in the legend). 
	Escape velocities range between~$418.4$~and~$423.4$~km/s, according to the model.}
\label{PDF_Outflows} 
\end{figure*}

Figure~\ref{PDF_Outflows} shows the histograms of the radial velocities of gas 
within different regions around the main halo of the four simulations 
{\sl BHs\_noFB} (blue), {\sl AGN\_~fid} (red), {\sl SF\_only} (light blue), and {\sl SF\_only\_lowFB} (orange), at $z=6$. 
We analyse inflowing and outflowing gas in terms of both gas mass fraction (top panels) 
and metal mass fraction (bottom panels). 
We consider the velocity distribution for all the gas in the computational volume (left panels), 
for the gas within the virial radius (middle panels), and 
for the gas within $0.1$~r$_{\rm vir}$ (right panels). 
We find that including quasar feedback results in gas outflowing at higher velocities 
within all the considered volumes, with respect to models that only account for stellar feedback.

\begin{table}
\centering
\begin{minipage}{85mm}
\caption[]{Outflowing gas mass fractions for different simulations ({\sl {Column~1}}) at $z=6$. 
{\sl {Column~2:}} escape velocity $v_{\rm esc}$ of the main subhalo. 
{\sl {Columns~3~and~4:}} fraction of outflowing gas mass with radial velocity exceeding $v_{\rm esc}$,
	within~r$_{\rm vir}$ and 0.1~r$_{\rm vir}$, respectively. 
{\sl {Columns~5~and~6:}} fraction of outflowing gas mass with radial velocity exceeding $1000 \text{ km/s}$,
	within~r$_{\rm vir}$ and 0.1~r$_{\rm vir}$, respectively.} 
\renewcommand\tabcolsep{1.8mm}
\begin{tabular}{@{}lccccc@{}}
\hline
Simulation  &  v$_{\rm esc}$   &   
\multicolumn{2}{c}{$\frac{\text{M}_{\rm gas} (\text{v}_{\rm rad}>\text{v}_{\rm esc})}{ \text{M}_{\rm gas} (\text{v}_{\rm rad}>0)}$}  &   
\multicolumn{2}{c}{$\frac{\text{M}_{\rm gas} (\text{v}_{\rm rad}>1000 \text{ km/s}) }{ \text{M}_{\rm gas} (\text{v}_{\rm rad}>0)}$}     \\ 
                  &                        &  r <  r$_{\rm vir}$   &  r < 0.1 r$_{\rm vir}$   &  r < r$_{\rm vir}$   &  r < 0.1 r$_{\rm vir}$   \\
                  &    (km / s)        & ($10^{-2}$)             &    ($10^{-2}$)             &      ($10^{-3}$)       &  ($10^{-3}$) \\
\hline
\hline
{\sl BHs\_noFB}              &  $421.3$  &   $2.53$ &    $3.01$   &   $0.65$   &  $1.51$ \\  
\hline
{\sl AGN\_~fid}                &  $423.4$  &  $3.79$  &    $3.16$  &  $1.58$   &  $1.7$ \\  
\hline
{\sl SF\_only}                  &  $418.4$  &   $3.78$  &   $2.24$  &  $0.25$   &  $0.2$  \\  
\hline
{\sl SF\_only\_lowFB}     &  $418.8$  &   $1.78$  &    $1.44$  &  $0.0$  &  $0.0$  \\  
\hline
\hline
\end{tabular}
\label{tab:GasOutflow}
\end{minipage}
\end{table}

Table~\ref{tab:GasOutflow} lists the outflowing gas mass fractions for different simulations. 
We quantify the mass of gas which is outflowing with radial velocity exceeding 
either the escape velocity of the halo or a reference threshold velocity of $1000$~km/s 
(over the total outflowing gas), considering gas within the virial radius and 0.1~r$_{\rm vir}$. 
The table summarises how remarkable the role of AGN feedback in driving outflowing gas to higher velocity is. 
We also find a larger outflowing gas mass fraction when the kinetic stellar feedback imparts 
higher velocities to gas surrounding SF sites (model {\sl SF\_only} wrt {\sl SF\_only\_lowFB}). 
However, assuming a stronger or weaker stellar feedback does not impact on outflowing gas velocity 
as significantly as the inclusion of quasar feedback does, especially when velocities above $1000$~km/s are considered. 

As a caveat, we note that fractions of outflowing gas mass with radial velocity larger than 
the local $v_{\rm esc}$ listed in Table~\ref{tab:GasOutflow} (columns $3$~and ~$4$) actually provide upper limits. 
Indeed, it cannot be excluded that gas initially moving at 
$v_{\rm rad} > v_{\rm esc}$ can eventually be slowed by 
ambient gas entrainment. 
We also note that velocity thresholds used to investigate outflow properties 
are relevant for final results.  While we adopt $v_{\rm rad} > 0$ to distinguish between 
outflowing and inflowing gas,  we acknowledge the possibility that a fraction of the gas orbiting 
in a deep potential well (as the one of our host systems) can reach $v_{\rm rad} \la 200$~km/s 
due to gravitational motions.  

As for gas inflow, we do not observe a significant difference between models {\sl AGN\_~fid} and {\sl SF\_only}. 
Indeed, we find that while being responsible for driving a larger amount of gas to outflow with larger speed, 
quasar feedback does not impact on inflowing ($v_{\rm rad} < 0$) gas. 
As different models have almost the same amount of (mainly cold) gas inflowing into the forming galaxy, 
they are experiencing a comparable accretion of gas from the large-scale environment. 
Hence, at $z=6$, quasar feedback is not yet capable of hampering the cosmological infall. 
This is also the main reason why we observe a similar SFH for the two aforementioned models, 
the amount of gas which fuels SF being still comparable down to $z=6$. 
We expect that gas infall from the large-scale structure will be halted at lower redshift, and that this will contribute 
to suppress the SF along with the additional activity of quasar feedback, whose long-term effect will be 
crucial at hindering gas accretion from outside and heating up the gas inside the galaxy 
(see also Section~\ref{sec:discussion}).

\section{Discussion} 
\label{sec:discussion}

\subsection{The role of quasar and stellar feedback} 
\label{sec:discussion_AGN}

The AGN activity resulting from our simulations can produce episodes of both negative and positive feedback: 
this finding is worth to be highlighted. 
Despite the negligible impact that AGN feedback has on the SFH of the host galaxy down to $z=6$, 
we find that the SF within $0.1$~r$_{\rm vir}$ in the reference model {\sl AGN\_~fid} not only 
can be suppressed, but also enhanced with respect to the simulation {\sl SF\_only} (Figure~\ref{fig:sfr}). 
Quasar feedback energy can suppress temporarily the SF because it heats up the gas (see e.g. 
Figures~\ref{Profiles}~and~\ref{Profiles_SFonly}). However, it can also enhance the SFR because it 
overpressurizes the star-forming gas\footnote{See \citet{Valentini2020} for details, 
	and for other similar evidence in previous numerical works.}. 
This result is also in line with \citet{Bischetti2020}, where an increased SF efficiency 
with respect to main sequence galaxies is observed in a sample of hyper-luminous quasars ($4 > z > 2$). 

Considering the small impact that quasar feedback has on final properties of its host galaxy 
(at $z=6$, on spatial scales of several pkpc), it can be interesting to investigate whether this result stems from 
the choice of the quasar feedback (and/or radiative) efficiency in our modelling. 
As already discussed in Section~\ref{BH_properties}, 
we adopted efficiency values to match BH masses on the $M_{\rm BH} - M_{\star}$ relation observed 
at high-redshift by \citet{Wang2010, Pensabene2020}. 
To quantify the relative importance of the SN and AGN feedback processes, we proceed as follows. 
The typical energy input per unit time injected by SN explosions within the virial radius 
in the model {\sl AGN\_~fid} at $z=6$ reads: 
\[
\dot{E}_{\ast} = (f_{\rm fb, therm} + f_{\rm fb, kin}) \, E_{\rm SN} \, \text{SFR} / M_{\star, \rm SN}  
\simeq 1.77 \cdot 10^{43} \text{ erg/s}\,\,,
\]
where approximate values of $\text{SFR} = 200$~M$_{\odot}$/yr and $M_{\star, \rm SN}=120$~M$_{\odot}$ 
have been assumed (see Section~\ref{sec:CSF} for further details). 
In the same simulation, AGN feedback supplies energy at the following rate: 
\[
\dot{E}_{\rm AGN} = \epsilon_{\rm f} \, \epsilon_{\rm r} \,  \dot{M}_{\rm BH}  \, c^2 
\simeq 0.61 \cdot 10^{43} \text{ erg/s}\,\,.
\]
Since $\dot{E}_{\rm AGN} / \dot{E}_{\ast} \simeq 0.35$, this explains why the effect of the quasar feedback is 
subdominant with respect to that of SNe at $z=6$. 
The relative contribution between AGN and stellar feedback increases when considering spatial scales smaller 
than r$_{\rm vir}$. In fact, in this case, $\dot{E}_{\ast}$ decreases as the SFR is lower (see Figure~\ref{fig:sfr}) 
while $\dot{E}_{\rm AGN}$ remains unchanged. 

The aforementioned estimate can be evaluated for all the other simulations that we carried out, 
by exploiting results listed in Table~\ref{tab:otherSims}. 
The impact of changing the quasar feedback efficiency can be appreciated for instance by comparing 
simulations {\sl AGN\_~fid} and {\sl AGN\_highFB}. We find that when a larger (by a factor of $10$) $\epsilon_{\rm f}$ 
is adopted, the impact of AGN feedback on the properties of the galaxy host (e.g. the distribution of gas particles 
in the density-temperature plane) is not significantly different with respect to the reference {\sl AGN\_~fid}. 
This can be explained by considering that $\dot{E}_{\rm AGN}$ depends linearly on 
both $\epsilon_{\rm f}$ and $\dot{M}_{\rm BH}$ 
(i.e. $\dot{E}_{\rm AGN} \propto \epsilon_{\rm f}$, while $\dot{E}_{\rm AGN} \propto M_{\rm BH}^2$), 
and that the $\dot{M}_{\rm BH}$ of the most massive BHs in the two simulations 
differ by a factor $\sim 400$ (see Table~\ref{tab:otherSims}). 
In conclusion, the result that quasar feedback does not affect significantly the final properties of the host galaxy 
does not depend on our choice of AGN feedback efficiencies, tuned to reproduce the $M_{\rm BH} - M_{\star}$ relation observed at high-redshift. 
Rather, our findings suggest that setting the SMBH on the observed $M_{\rm BH} - M_{\star}$ relation 
implies that its feedback is subdominant with respect to stellar feedback. 

The lack of SF quenching in the simulation {\sl AGN\_~fid} with respect to {\sl SF\_only} 
does not exclude that SF can be shut down at $z<6$. 
A more cumulative and long-term impact of AGN feedback on the host galaxy can later suppress SF. 

The way in which AGN feedback is numerically implemented in our code contributes to determine 
the results discussed so far. 
Kinetic energy deposition, not included in this work, might be an important addition to the thermal one considered here.  
Since the kinetic injection of AGN feedback energy is expected to produce stronger signatures (kinetic energy 
thermalising by construction later and at larger scales; see Section~\ref{sec:introduction}), we predict that simulations 
adopting only a mechanical AGN feedback have a significantly higher impact on the host galaxy. 
We envisage that in a hybrid scenario where thermal and mechanical AGN feedback act in tandem to shape 
BH and galaxy evolution, the kinetic feedback is crucial to eject gas from the innermost regions of forming structures, 
thus reducing the surrounding gas column density and contributing to quench SF. 
We note that it is not straightforward to numerically achieve the joint activity of thermal and kinetic AGN feedback 
in cosmological simulations: for instance, accurate hybrid models \citep[e.g.][]{Weinberger2017} as the one adopted 
in the Illustris-TNG simulations \citep{Pillepich2018} which consider the BH accretion rate to discriminate whether the 
feedback has to be thermal or kinetic, would result in a thermal AGN feedback only with accretion rates 
characterising quasars (see e.g. Figure~\ref{BHmdot}). 
Another possible direction of investigation and improvement is represented by the way in which 
AGN feedback energy is provided to the gas surrounding the BH. In fact, as the resolution of simulations increases, 
the resolution elements around the BH which are provided with AGN feedback energy occupy an always smaller region. 
The reduced volume where feedback energy is injected can play a role in determing to what extent the AGN feedback 
is effective, and the farthest scale affected by the process. 
The investigation of these effects is postponed to a forthcoming work. 
We also postpone to an upcoming study a detailed analysis of inflow and outflow rates, with the final goal of 
comparing predictions from our simulations to available estimates from observations. 

As for the expected number density of UV low-luminosity quasars, the intrinsic (dust unabsorbed) UV magnitude of the most massive BH in the simulation 
{\sl AGN\_~fid} is $M_{\rm UV} = - 25.6$ at $z=6$ (Section~\ref{BHmasses}); we expect a corresponding observed (dust extinguished) UV magnitude $M_{\rm UV, obs} \simeq - 24$, at $z=6$ \citep{DiMascia2021b}. 
Quasars of this magnitude correspond to the low-luminosity tail explored by \citet{Matsuoka2016}, and to a number density of $\sim 10^{-8}$~Mpc$^{-3}$ at $z=6$. The latter number density exceeds the number density of haloes with $\sim 10^{12}$~M$_{\odot}$ at $z=6$ (as in our suite of simulations) by a factor of $\sim 10^2$ \citep[e.g.][]{Angulo2012}.  

This result does not imply that our SMBH growth and feedback model overestimates the number density of $z\sim 6$ quasars, since the aforementioned numbers can be reconciled either assuming that the duty-cycle ($\tau_{DC}$) of SMBHs hosted in $\sim 10^{12}$~M$_{\odot}$ haloes is short ($\sim 10^{-2}$) at $z = 6$, or that our {\sl AGN\_~fid} run can be considered representative only of one out of $\sim 100$ of them.  

Figure~\ref{BHmdot} shows that $\tau_{DC}\sim 1$ for {\sl AGN\_~fid}, for values of the Eddington-ratio ($\lambda_{\rm Edd}=\dot{M}_{\rm BH} / \dot{M}_{\rm Edd}\sim 0.01-0.1$) typically adopted to distinguish between on/off AGN activity \citep[e.g.][]{Delvecchio2020} or high-/low-accretion phases (see Section~\ref{BHmdot}). 
As a consequence, the first hypothesis is unlikely, unless $\lambda_{\rm Edd}\simeq 1$ is considered as the threshold to asses whether the quasar is active.  

The second hypotesis is instead supported by the results reported in Table~\ref{tab:otherSims}: haloes with $\sim 10^{12}$~M$_{\odot}$ at $z=6$ do not necessarily always host quasars as luminous as that in our {\sl AGN\_~fid} model. Our simulation {\sl AGN\_~fid} has been designed to investigate the evolution of a BH which grows supermassive (to~$\sim 10^9$~M$_{\odot}$) by $z=6$ in a $\sim 10^{12}$~M$_{\odot}$ halo. To this goal, BH radiative and feedback efficiencies have been tuned to the adopted values 
(Sections~\ref{sec:AGN},~\ref{sec:simset}~and~\ref{BHmago}).

\subsection{Comparison with observations}
\label{comp_obs}

Several observations suggest the presence of SFR- and/or AGN-driven outflows within the ISM of galaxies and AGN, at low and high redshift. However, no striking differences have been so far outlined between different systems, also because of the loosely constraining, available data. Here, we revise the most recent observational results in normal star-forming galaxies and AGN, and compare them with predictions from our simulations.

ALMA observations of high-redshift ($5<z<6$), normal star-forming galaxies 
($\text{SFR}=10-100$~M$_{\odot}$~yr$^{-1}$) show broad [CII] wings, suggestive of cold, neutral gas outflowing with velocity up to $\sim 500$~km~s$^{-1}$ \citep[e.g.][]{Gallerani2018, Sugahara2019, Ginolfi2020}. These results are not dissimilar from the ones inferred from local observations of SF-driven outflows, shown to correlate with the SFR, and to have typical velocity 
spanning the range $300-800$~km~s$^{-1}$ in galaxies with SFR as high as~$\sim200$~M$_{\odot}$~yr$^{-1}$ \citep[e.g.][]{ForsterSchreiber2019}. Also, \citet{Martin2005} analyses large-scale outflows in a sample of SF-dominated ultraluminous galaxies and finds that the upper limit of the outflow velocity 
of the warm, neutral (T$\leq 10^4$~K) gas is $\sim 400-500$~km~s$^{-1}$, 
with quite a large scatter towards lower values. Moreover, \citet{Heckman2015} investigate far-UV absorption lines for low-redshift, starburst galaxies (with physical properties 
akin to those of high-redshift, Lyman Break galaxies) and infer a velocity of $\sim350-650$~km~s$^{-1}$ for the 
warm ionized phase of starburst-driven winds in galaxies having a SFR of~$\sim200$~M$_{\odot}$~yr$^{-1}$.
Finally, \citet{Cicone2016} find that the ionised gas is outflowing at $\sim 600-800$~km~s$^{-1}$ 
in a large sample of normal, star-forming galaxies in the Sloan Digital Sky Survey. 

Our {\sl SF\_only} simulation predicts outflows with velocities up to $\sim 500~\rm km~s^{-1}$ driven by $\text{SFR}=200~\rm M_{\odot}yr^{-1}$, consistently with the aforementioned results. 

AGN observations seem to suggest that the presence of an active BH increases the maximum speed that galactic outflows can reach. This is in line with our result that quasar feedback is more effective than stellar feedback at driving gas to the largest outflow velocities found in our simulations ($v\sim 1500~\rm km~s^{-1}$). Outflow velocities can be as high as $\sim 700$~km~s$^{-1}$ in optically 
selected quasars at $z \sim 6$ \citep{Stanley2019}, or even more extreme ($\sim 1000-1500$~km~s$^{-1}$), as in the case of J1148+5251 at $z=6.4$ \citep[][but see also \citealt{Decarli2018,novak2020}]{Maiolino2012, Cicone2015}.

At lower redshift, \citet{Fiore2017} study the connection between extended AGN winds 
and host galaxy properties in a sample of AGN, including hyper-luminous quasars at $2 < z < 3$. 
For systems having a SFR of~$\sim 200$~M$_{\odot}$~yr$^{-1}$ 
(although their SFR is not actually the instantaneous SFR as in our simulations) 
and a typical AGN bolometric luminosity of~$10^{46}-5\cdot10^{47}$~erg~s$^{-1}$, 
they find that the maximum ionised wind velocity can be in the range $500-3000$~km~s$^{-1}$. \citet{cicone2014} investigate galactic-scale, molecular outflows in a sample of local galaxies characterised by different AGN and starburst activity and conclude that 
even if the AGN does not represent the dominant source of energy (see Section~\ref{sec:discussion} for our models), 
still it can be more effective at promoting outflows than SF activity. 

Overall, observations loosely constrain outflow velocity and often suggest expected velocity ranges at redshift intervals 
which can be different from the one we have focussed on in this work, data availability being larger in the local universe. 
The general agreement between our results and observations is remarkable, as it is the trend for outflow velocity 
to be higher when AGN drives winds in addition to SF activity. 
The comparison between predictions from simulations and observations is not straightforward for 
a few reasons: for instance, the spatial scale probed by observations often cannot be certainly established. 
Moreover, the issue of fairly comparing gas phases probed in simulations with those traced by different observables 
is not a trivial one. Indeed, different phases within the same resolution element are forced to move together 
in simulations, and thus it is not possible to take into account the case where phase coupling is not present or 
hot gas entrainment by the cold phase is not achieved.

\subsection{Comparison with other numerical works} 
\label{sec:comp_sims}

The BH accretion model that we adopt allows us to form SMBHs with masses in agreement 
with the observed $M_{\rm BH} - M_{\star}$ relation at $z=6$, without the need of assuming a boost factor 
and by even suppressing the accretion of cold gas with high angular momentum 
(thus improving the commonly adopted Bondi model, see Section~\ref{sec:AGN}). 
This also highlights how the AGN feeding and feedback processes are tightly linked in our simulations. 
In our model, the amount of quasar feedback energy coupled to the ISM which surrounds the BH 
determines directly the properties of the gas which is later 
accreted onto the BH and which are used to compute the BH accretion rate. 
Larger AGN feedback efficiencies would couple a larger amount of energy to the ambient gas: 
as a consequence, the gas is heated, its density decreases, 
and so does the BH accretion rate (equation~(\ref{BondiMdotAve})). 

This is a key difference with respect to models that adopt a boost factor to describe the AGN feeding process 
(e.g. \citealt{Springel2005e361, DiMatteo2005, Sijacki2007, Khalatyan2008, BoothSchaye2009, Dubois2013, 
Costa2014MNRAS439, Barai2018}; but see \citealt{Lupi2019}). 
In those models, a way larger amount of AGN feedback energy can be coupled to the gas around the BH 
without affecting directly the BH accretion process (whereas more evident feedback signatures on the host galaxy 
can be produced). Indeed, if AGN feedback heated the gas and decreased its 
density, the presence of a fudge factor (whose value tipically ranges from several tens to few hundreds) 
in those models would compensate for a low BH accretion rate. 
Since almost all the simulations adopt the $M_{\rm BH} - M_{\star}$ relation to get BH final masses in 
agreement with observations, this explains why the quasar feedback efficiency adopted 
in the reference model {\sl AGN\_~fid} is lower than commonly assumed. 

Our simulation suite provides a detailed outlook on the processes of SMBH growth, quasar feedback 
and outflows in the early universe. 
Our finding that BHs can grow supermassive ($\sim 10^8 - 10^9$~M$_{\odot}$) 
from massive ($\sim 10^5 - 10^6$~M$_{\odot}$) seeds 
in massive ($\sim 10^{12}$~M$_{\odot}$) DM haloes by $z=6$ via Eddington-limited gas accretion 
is consistent with results from several, previous simulations
\citep[e.g.][]{Sijacki2009, dimatteo2012, dimatteo2017, Costa2014MNRAS439,
Smidt2018, Barai2018, Lupi2019, Zhu2020}.
In line with our results, works among the aforementioned ones have also shown 
that SMBHs often accrete at a rate which is close to the Eddington rate 
\citep[e.g.][]{dimatteo2017, Smidt2018, Barai2018, Lupi2019}, 
and BH-BH mergers contribute little to the BH growth with respect to gas accretion 
\citep[][]{dimatteo2012, dimatteo2017}. 
The numerical modelling of processes driving (or hampering) the BH growth features 
differences between our work and previous simulations, and among the aforementioned works themselves.  

Powerful outflows in our simulations are triggered by the joint activity of stellar and quasar feedback, 
in agreement with results by \citet{Costa2015, Biernacki2018},  who also showed that the AGN-powered 
component is necessary for the outflows to reach higher radial velocities.  
Our finding that cold gas in outflows moves with a slower speed than the hot phase component 
is in line with previous works; however,  cold gas outflows in our simulations are slower 
than those predicted by \citet{Costa2015, Ni2018}, both the aforementioned works 
retrieving for them velocities which can even exceed $\sim 1000$~km/s.  

As for cold gas inflow,  our finding that inflowing warm and cold gas filaments feed the halo 
and provide the growing BH and the forming host galaxy with fuel is in agreement with results from previous 
simulations \citep{Sijacki2009, dimatteo2012, Dubois2013, Costa2014MNRAS439, Barai2018, Smidt2018}.
While there is general consensus on the role played by these cold streams in funnelling gas towards 
the innermost regions of growing structures, it is still debated whether the complex web of filaments 
can survive the effect of AGN feedback due to their high density \citep{dimatteo2012}, 
although dynamically perturbed \citep{Dubois2013}, 
or can be disrupted by quasar outflows propagating in the same direction \citep{Barai2018}.  
Our simulations support the idea that inflowing cold gas streams cannot be halted 
by the joint SN and quasar feedback by $z=6$.  

As also discussed in previous sections, quasar feedback controls the SMBH growth 
\citep[see also][]{Sijacki2009, Dubois2013} and affects to various degrees host galaxy properties 
\citep[e.g.][]{dimatteo2012, Khandai2012, Costa2014MNRAS439, CurtisSijacki2016, Habouzit2019}. 
Interestingly,  quasar feedback in the aforementioned simulations is found to suppress SF little to moderatly, 
the full quenching being never achieved by $z \sim 6$.  
A longer-term AGN feedback effect is expected to be crucial to significantly suppress and even quench SF at lower redshift, 
as we envisage in Sections~\ref{outflows_res}~and~\ref{sec:discussion_AGN} 
\citep[see also][]{CurtisSijacki2016}.  

Although the common expectation is that the inclusion of AGN feedback results in striking differences 
with respect to the case where SMBH effects are not accounted for, some recent works taking 
advantage of state-of-the-art cosmological simulations showed that this may be not always true. 
Recently, \citet{Sorini2020} investigated the properties of the circumgalactic and intergalactic medium around 
quasars at redshift $2 < z < 3$ in the SIMBA simulation. Interestingly, they found that the physical properties 
of the gas surrounding quasars, i.e. gas density, temperature, and radial velocity out to several virial radii, are 
primarily shaped by stellar feedback, while the contribution from the mechanical AGN feedback 
(in different flavours, namely winds, jets and X-ray heating) plays a minimal role. 
Similar conclusions have been also drawn by \citet{Rahmati2015}: when analysing the distribution of neutral 
hydrogen around high-redshift ($2 < z < 3$) galaxies and quasars in the EAGLE simulation, they found that 
the neutral hydrogen covering fraction in Lyman Limit Systems is not sensitive to the effect of AGN feedback 
at all (out to $\sim 1$~pMpc), while the stellar feedback is the main driver for the results. 
Results from \citet{FaucherGiguere2016} are also in line with the finding that the availability of neutral hydrogen 
(on $\sim 100$~kpc scale) 
is mainly determined by the effect of stellar feedback alone: by studying properties of massive haloes 
(10$^{12} < {\text {M}}_{\rm halo} ({\text {M}}_{\odot}) < 10^{13}$ at $z=2 - 2.5$) within the FIRE project,
they found neutral hydrogen covering fractions in agreement with observations of luminous quasars and claim 
that a significant contribution from AGN feedback is not needed. 
In addition to the aforementioned numerical studies, recent observations \citep[e.g.][]{Davies2020, Scholtz2020} 
of galaxies at $z \lesssim 2.5$ have also supported the evidence that AGN outflows can have no effect on 
the instantaneous SFR of the host galaxy.

\section{Summary and Conclusions}
\label{sec:conclusions}

We carried out a suite of high-resolution 
($m_{\rm gas} = 2.89 \cdot 10^5$~M$_{\odot}$ and $\epsilon = 59$~ppc for gas particles) 
cosmological, zoom-in simulations of high-z galaxies 
($M_{\rm halo, \, DM} \simeq 10^{12}$~M$_{\odot}$, at $z=6$), based on the GADGET-3 code 
and using the MUPPI sub-resolution model to describe physical processes in a multiphase ISM, 
BH accretion and thermal quasar feedback. 
The goal of this study is to investigate the growth history of SMBHs down to $z=6$, 
and to quantify the impact of stellar and quasar feedback both on the quasar-host galaxy final properties 
and on the formation of SMBHs. 
Our main results can be summarised as follows:
\begin{itemize}
\item BHs can grow supermassive by $z=6$ and reach a final mass which is in agreement with the $M_{\rm BH} - M_{\star}$ relation observed at that redshift (Figure~\ref{Magorrian}). Gas accretion is the main driver of BH growth, with mergers playing a sub-dominant role (Figure~\ref{BHmassGrowth}). 
\item In our reference model, {\sl AGN\_~fid}, the central, most massive BH has a mass of $9.85 \cdot 10^8$~M$_{\odot}$, an accretion rate $\dot{M}_{\rm BH}=35.53$~M$_{\odot}$~yr$^{-1}$ (i.e.  $\dot{M}_{\rm BH} / \dot{M}_{\rm Edd} = 0.495$), and is hosted in a galaxy whose stellar mass is $M_{\ast} \sim 2.6 \cdot 10^{10}$~M$_{\odot}$, at $z=6$. 
Such $\dot{M}_{\rm BH}$ value corresponds to an intrinsic, unextincted 
UV magnitude of $M_{\rm UV} = - 25.6$.  
If the quasar feedback efficiency were tuned to produce a SMBH 
lying on the \textit{local} $M_{\rm BH} - M_{\star}$ relation, 
its accretion rate at $z=6$ would fall short of the measured one in high-$z$ quasars. 
\item The slope of the $M_{\rm BH} - M_{\star}$ relation inferred from {\sl AGN\_~fid} is steeper than suggested by local observations. At high redshift, the $M_{\rm BH} - M_{\star}$ relation in our models is shaped by quasar feedback, 
which controls the BH growth while leaving SF almost unaffected (due to its inability to halt the cosmological infall). 
\item By comparing properties of the ISM in models with and without SMBH, we find that 
the temperature of the ISM is on average higher (by a factor of~$\sim 2$ within $\sim 4$~kpc from the galaxy centre) and that the total gas metallicity is lower (by a factor of~$3$) 
due to a reduced SFR (by~$10$~M$_{\odot}$~yr$^{-1}$) when AGN is not included. Properties of the host galaxy in our fiducial simulation are in good agreement with observations. 
\item Quasar feedback has two opposite effects on the SFH of the host galaxy: (a) by heating the gas, it quenches SF; (b) by overpressurizing the ISM, it favours the formation of new stars. However, such modulation effects are sub-dominant with respect to the rate imposed by cosmological infall (see below). As a result, feedback has only a negligible effect on the galaxy SFH. Nevertheless, quasar feedback strongly controls BH growth. When turned off in simulations, the final SMBH mass is found to be $\approx 100\times$ larger. 
\item Galactic outflows are promoted by the joint activity of stellar and quasar feedback. We find that 
quasar feedback increases the outflow rate and accelerates the gas to larger velocities (Figure~\ref{PDF_Outflows}).  
Hot and cold phases are both involved in outflows. In addition, different phases are characterised by different kinematics (Figure~\ref{Outflows}): 
the hot ($T \gtrsim 10^5$~K) gas has velocities which can easily exceed the escape velocity of the halo 
and be even larger than $\sim 1000$~km/s, when quasar feedback is included. 
On the other hand, cold and warm ($T \lesssim 10^4$~K) phases have lower velocities, 
only in a few cases exceeding $\sim 300$~km/s. 
The imprint of quasar feedback is on the high-velocity tail of the outflowing gas distribution; this feature is present even if the AGN does not represent the dominant source of energy in the host galaxy. 
Predictions from our simulations as for outflow velocity are in good agreement with observations. 
\item Cold gas makes up for the almost totality of the infalling mass. 
We find that quasar feedback cannot hinder the inflow process. Models with and without 
SMBH activity experience a comparable accretion rate from the large-scale structure; such  
cosmological infall fuels SF at a comparable rate in different systems. 
\end{itemize}

The suite of simulations introduced in this work will be further analysed in forthcoming papers. 
As new upcoming observational instruments (e.g. the James Webb Space Telescope, JWST and the
European-Extremely Large Telescope, E-ELT) will allow to probe the very high-redshift Universe, it is extremely important to have simulations able to provide the theoretical counterpart and to shed
light on what drives the formation and evolution of the first structures.


\appendix


\section{Haloes in the parent, DM-only simulation}
\label{DMsim}

\begin{figure*}
\newcommand{\captionfonts}{\small}
\begin{minipage}{\linewidth}
\centering
\includegraphics[trim=1.5cm 0.cm 0.cm 1.cm, clip, width=1.\textwidth]{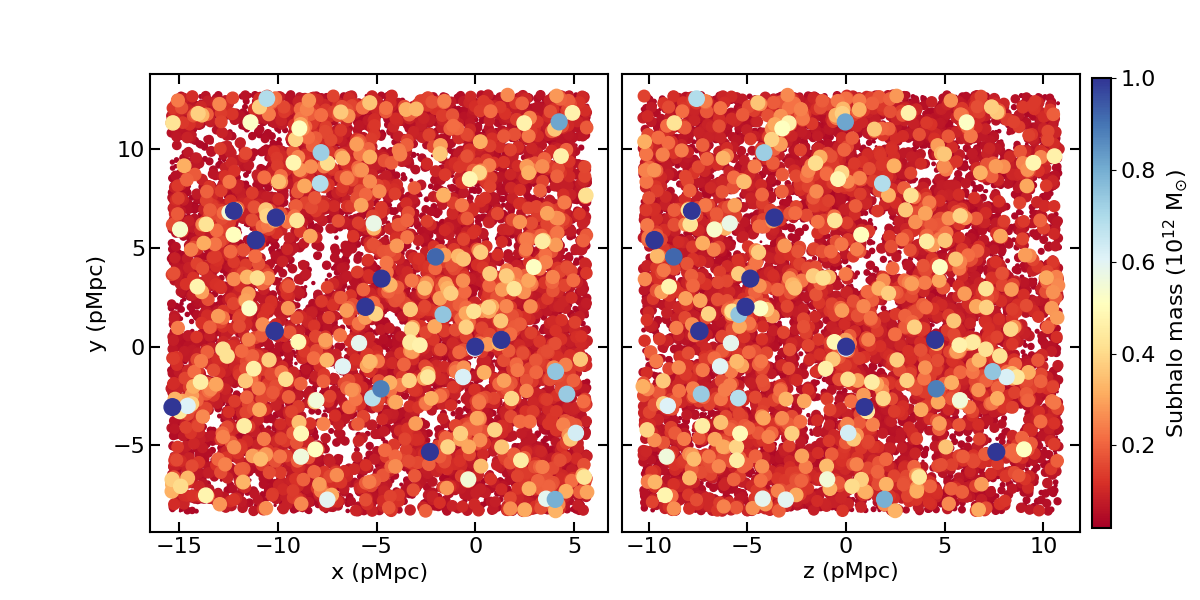} 
\end{minipage} 
\caption{Projected distribution of all the subhaloes in the parent, DM-only simulation 
	in the planes $x-y$ {\sl (left-hand panel)} and $z-y$ {\sl (right-hand panel)}, at redshift $z = 6$. 
	The colour bar encodes the subhalo mass. 
	The size of each circle scales with the virial radius of the subhalo. 
	Distances are expressed in physical Mpc (pMpc), 
	with respect to the centre of the subhalo that has been chosen for the zoomed simulation.}
\label{HaloMaps} 
\end{figure*}

\begin{figure}
\newcommand{\captionfonts}{\small}
\centering
\includegraphics[trim=0.2cm 0.cm 0.5cm 0.5cm, clip, width=.5\textwidth]{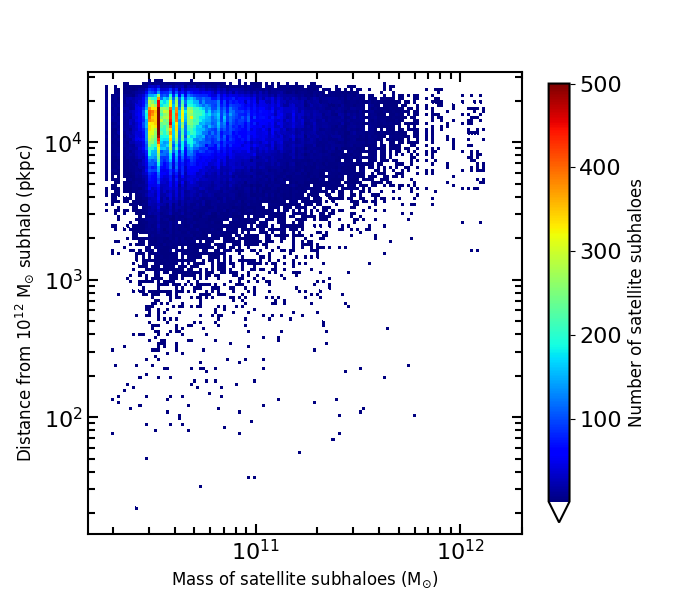}  
\caption[]{Cumulative 2-D histogram providing the number of satellite subhaloes per bin of mass and distance 
	from each of the subhaloes whose mass is larger than $10^{12}$~M$_{\odot}$. 
	This histogram has been obtained by summing up the histograms of each of the $10$ subhaloes 
	more massive than $10^{12}$~M$_{\odot}$ in the parent DM-only simulation, at $z=6$. 
	The colour in each bin encodes the number of subhaloes.}
\label{HaloHist} 
\end{figure}

In this Appendix, we highlight some interesting features of the parent, DM-only simulation 
described in Section~\ref{sec:ICs}, and discuss how we selected our target halo for the zoomed-in simulation. 

Figure~\ref{HaloMaps} shows the distribution of all the subhaloes in the parent, DM-only simulation 
identified by the \textsc{SUBFIND} algorithm at $z = 6$. The left-hand and the right-hand panel 
depict the $x-y$ and $z-y$ projections, respectively. 
The colour of each circle encodes the subhalo mass, while its size is proportional to the virial radius of the subhalo. 
Distances are shown with respect to the centre of the target subhalo, located in $(x,y,z) = (0,0,0)$. 
As for the statistics of the subhaloes, 
there are $10$ subhaloes more massive than $10^{12}$~M$_{\odot}$ (blue circles), 
$41$ subhaloes whose mass exceeds $5 \cdot 10^{11}$~M$_{\odot}$, 
$1892$ ($6584$) subhaloes more massive than $10^{11}$~M$_{\odot}$ ($5 \cdot 10^{10}$~M$_{\odot}$). 
The most massive subhalo in the box has a mass of $1.32 \cdot 10^{12}$~M$_{\odot}$, 
the target halo is as massive as $1.12 \cdot 10^{12}$~M$_{\odot}$ (ranked $8$th). 

When selecting the target subhalo, we excluded subhaloes with mass~$> 10^{12}$~M$_{\odot}$ close to the 
borders of the box. 
We also excluded massive~($> 10^{12}$~M$_{\odot}$) subhaloes in underdense regions, 
i.e. with no close, fairly-massive subhaloes, as possible candidate target subhaloes. 
For instance, the subhaloes ranked $2$nd, $7$th, and $9$th have been excluded because there are no 
subhaloes at least as massive as~$8 \cdot 10^{10}$~M$_{\odot}$ within a distance of~$6.5$~times 
their own virial radius. 
In addition, we decided not to focus on a subhalo more massive than $10^{12}$~M$_{\odot}$ 
with a too-close massive companion. 
For instance, we excluded the most massive subhalo as possible target because 
a~$\sim 2.6 \cdot 10^{11}$~M$_{\odot}$ subhalo is located at a distance which is smaller than the sum of the 
virial radii of the two aforementioned subhaloes. 
We selected our target halo because it is massive enough to be eligible for a quasar-host galaxy, and 
because it has a massive ($\sim 7.5 \cdot 10^{11}$~M$_{\odot}$) satellite subhalo 
$\sim 103$~pkpc far from it (this distance being larger than the sum of the virial radii of the two systems). 

The proximity between massive subhaloes is indeed a very interesting topic, 
as several studies have suggested 
that extremely-massive BHs ($M_{\rm BH} \gtrsim 10^8 - 10^9$~M$_{\odot}$) 
preferably reside in overdense regions \citep[see e.g.][and references therein]{Yoon2019}. 
Thus, we included the distance from a massive, satellite subhalo as a requirement to select the target halo. 
A close (non merging, see below), massive subhalo can indeed shed some light on the joint evolution of 
the host BHs and on their environment. This subject will be further investigated in forthcoming works, too. 

To this end, Figure~\ref{HaloHist} shows the cumulative histogram of satellite subhaloes 
per bin of mass and distance from each of the $10$ most massive subhaloes, 
in the parent DM-only simulation, at $z=6$. 
We computed the histogram of the subhaloes surrounding each of the $10$ subhaloes 
more massive than $10^{12}$~M$_{\odot}$ as a function of both their mass and distance, and then 
summed up to retrieve a more solid result. 
The virial radius of the $10$ most massive subhaloes spans the range $46.9 - 50.7$~pkpc: 
as a consequence, if two subhaloes are closer than $\sim 10^2$~pkpc, they are likely interacting and 
will merge.
Figure~\ref{HaloHist} suggests how hard it is to have close, massive subhaloes, 
in the volume considered by our simulation. 
As these close, massive subhaloes are expected to be the hosts of extremely-massive BHs, 
this result hints at the unlikely possibility of having extremely-massive BH pairs in 
overdense regions of the early universe. Rather, massive subhaloes (and hence extremely-massive BHs) 
tend to reside in relatively isolated environments.


\section{Other simulations}
\label{OtherSims}

\begin{table*}
\centering
\caption[]{Relevant parameters adopted and main features of the simulated structures at $z=6$ in the suite of simulations carried out. 
{\sl {Column~1:}} simulation. 
{\sl {Column~2:}} kinetic stellar feedback efficiency. 
{\sl {Column~3:}} BH radiative efficiency. 
{\sl {Column~4:}} quasar feedback efficiency. 
{\sl {Column~5:}} SF efficiency. 
{\sl {Column~6:}} pressure of the ISM at which $f_{\rm mol}=0.5$ (see Section~\ref{sec:CSF}).
{\sl {Column~7:}} central BH mass. 
{\sl {Column~8:}} BH accretion rate. 
{\sl {Column~9:}} stellar mass within the virial radius. 
{\sl {Column~10:}} SFR within r$_{\rm vir}$. 
{\sl {Column~11:}} stellar mass within $0.1$~r$_{\rm vir}$. 
{\sl {Column~12:}} SFR within $0.1$~r$_{\rm vir}$. } 
\renewcommand\tabcolsep{1.4mm}
\begin{tabular}{@{}lcccccccccccc@{}}
\hline
\multicolumn{2}{c}{Simulation}      &  $f_{\rm fb, kin}$ & $\epsilon_{\rm r}$ & $\epsilon_{\rm f}$ &  $f_{\star}$  &	$P_0$  & $M_{\rm BH}$     & $\dot{M}_{\rm BH}$    & $M_{\ast}{\small{(\text{< r}_{\rm vir})}}$  & SFR${\small{(\text{< r}_{\rm vir})}}$ & $M_{\ast}{\small{(\text{< 0.1 r}_{\rm vir})}}$  & SFR${\small{(\text{< 0.1 r}_{\rm vir})}}$\\ 
                 &                                  &           &               &                   &                       &  (k$_{\rm B}$K/cm$^3$)  &   (M$_{\odot}$)      &  (M$_{\odot}$ / yr)   & ($10^{10}$~M$_{\odot}$)                  & (M$_{\odot}$ / yr)       & ($10^{10}$~M$_{\odot}$)                  & (M$_{\odot}$ / yr)\\
\hline
\hline
{\sl AGN\_~fid}                &  $1 $  &  $ 0.12$  &  $0.03$  &  $10^{-4}  $  &  $0.06$  &  $2 \cdot 10^4$  &  $9.85\cdot 10^8$  &  $35.53$  &  $4.03$  &  $205$  &  $1.13$  &  $80$\\  
\hline
{\sl BHs\_noFB}              &  $2 $  &  $0.12$  &  $0.0$  &  $0.0  $  &  $0.06$  &  $2 \cdot 10^4$  &  $4.62\cdot 10^{11}$  &  $3.17\cdot 10^4$  &  $3.51$  &  $190$  &  $0.68$  &  $75$\\  
\hline
{\sl AGN\_highFB}              &  $3 $  &  $0.12$  &  $0.03$  &  $10^{-3}  $  &  $0.06$  &  $2 \cdot 10^4$  &  $4.16\cdot 10^7$  &  $8.53\cdot 10^{-2}$  &  $4.03$  &  $210$  &  $1.11$  &  $85$\\  
\hline
{\sl SF\_only}              &  $4 $  &  $0.12$  &  $ - $  &  $ -$  &  $0.06$  &  $2 \cdot 10^4$  & $ - $   &  $ -$  &  $4.01$  &  $215$  &  $1.07$  &  $90$\\  
\hline
{\sl SF\_only\_lowFB}              &  $5 $  &  $0.05$  &  $ -$  &  $   -$  &  $0.06$  &  $2 \cdot 10^4$  &  $  -$  &  $   -$  &  $4.55$  &  $255$  &  $1.44$  &  $110$\\  
\hline
  &  $6 $  &  $0.12$  &  $0.1$  &  $0.01 $  &  $0.02$  &  $2 \cdot 10^4$  &  $7.26\cdot 10^6$  &  $1.06\cdot 10^{-2}$  &  $1.66$  &  $85$  &  $0.43$  &  $30$\\  
\hline
  &  $7 $  &  $ 0.12$  &  $0.05$  &  $0.01 $  &  $0.02$  &  $2 \cdot 10^4$  &  $1.05\cdot 10^7$  &  $2.37\cdot 10^{-2}$  &  $1.71$  &  $95$  &  $0.51$  &  $45$\\  
\hline
  &  $8 $  &  $0.12$  &  $0.1$  &  $2 \cdot 10^{-3} $  &  $0.02$  &  $2 \cdot 10^4$  &  $1.01\cdot 10^7$  &  $2.72\cdot 10^{-2}$  &  $1.72$  &  $90$  &  $0.54$  &  $40$\\  
\hline
  &  $9 $  &  $0.12$  &  $0.02$  &  $10^{-4} $  &  $0.02$  &  $2 \cdot 10^4$  &  $1.17\cdot 10^{10}$  &  $57.49$  &  $1.65$  &  $75$  &  $0.42$  &  $30$\\  
\hline
  &  $10 $  &  $0.12$  &  $0.1$  &  $5 \cdot 10^{-4} $  &  $0.02$  &  $2 \cdot 10^4$  &  $2.87\cdot 10^7$  &  $0.1 $  &  $1.71$  &  $85$  &  $0.49$  &  $35$\\  
\hline
  &  $11 $  &  $0.12$  &  $0.1$  &  $0.01$  &  $0.02$  &  $4 \cdot 10^3$  &  $6.5 \cdot 10^6$  &  $1.32\cdot 10^{-2}$  &  $2.03$  &  $90$  &  $0.49$  &  $35$\\  
\hline
  &  $12 $  &  $0.12$  &  $0.1$  &  $0.01 $  &  $0.02$  &  $30$  &  $5.51\cdot 10^6$  &  $2.81 \cdot 10^{-3}$  &  $2.54$  &  $100$  &  $0.49$  &  $35$\\  
\hline
  &  $13 $  &  $0.12$  &  $0.0$  &  $0.0  $  &  $0.02$  &  $2 \cdot 10^4$  &  $2.74 \cdot 10^9$  &  $59.49$  &  $1.7$  &  $80$  &  $0.52$  &  $40$\\  
\hline
  &  $14 $  &  $0.12$  &  $ -$  &  $  - $  &  $0.02$  &  $2 \cdot 10^4$  &  $ -$  &  $ -$  &  $1.81$  &  $80$  &  $0.47$  &  $30$\\  
\hline
  &  $15 $  &  $0.05$  &  $0.03$  &  $10^{-3} $  &  $0.02$  &  $2 \cdot 10^4$  &  $2.34 \cdot 10^8$  &  $5.84$  &  $2.14$  &  $130$  &  $0.73$  &  $70$\\  
\hline
  &  $16 $  &  $0.05$  &  $0.02$  &  $5 \cdot 10^{-4} $  &  $0.02$  &  $2 \cdot 10^4$  &  $9.25 \cdot 10^8$  &  $2.91$  &  $2.07$  &  $100$  &  $0.64$  &  $40$\\  
\hline
\hline
\end{tabular}
\label{tab:otherSims}
\end{table*}

\begin{figure*}
\newcommand{\captionfonts}{\small}
\begin{minipage}{\linewidth}
\centering
\includegraphics[trim=0.cm 0.cm 0.cm 0.cm, clip, width=1.\textwidth]{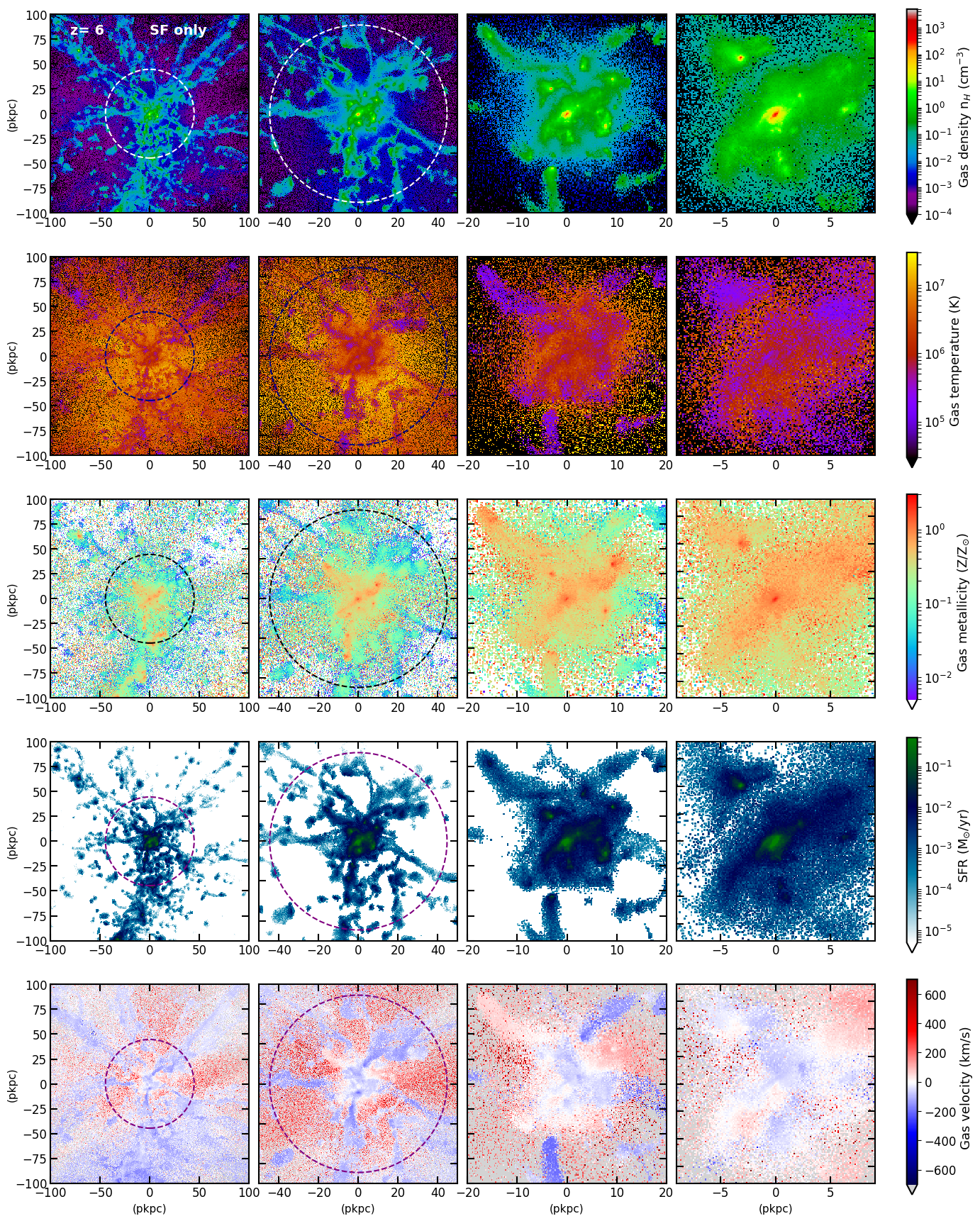} 
\end{minipage} 
\caption{We show gas density ({\sl first row}),
	gas temperature ({\sl second row}), 
	gas metallicity ({\sl third row}),
	the SFR of gas particles ({\sl fourth row}), 
	and the mass-weighted, radial velocity of gas particles ({\sl bottom row}) 
	for the simulation {\sl SF\_only}, at redshift $z = 6$. 
	We progressively zoom-in from left to right: 
	the {\sl first and second columns} show a box of $200$~pkpc and $100$~pkpc a side, respectively, 
	the projection being performed along the $z$-axis (over $200$~pkpc and $100$~pkpc, respectively). 
	The dashed circumference has the virial radius of the central, target halo as a radius. 
	The {\sl third column} shows a box of $40$~pkpc (projection is over $20$~pkpc along the $z$-axis), 
	while in the {\sl fourth column} we consider a box of $18$~pkpc 
	(projection is over $9$~pkpc along the $z$-axis). 
	All the maps are centred on the centre of the most massive subhalo.
	Same as Figure~\ref{IntroMapsZoom}, but for the simulation {\sl SF\_only}.}
\label{IntroMapsZoom_SFonly} 
\end{figure*}

\begin{figure*}
\newcommand{\captionfonts}{\small}
\begin{minipage}{\linewidth}
\centering
\includegraphics[trim=1.8cm 1.5cm 3.5cm 2.5cm, clip, width=.49\textwidth]{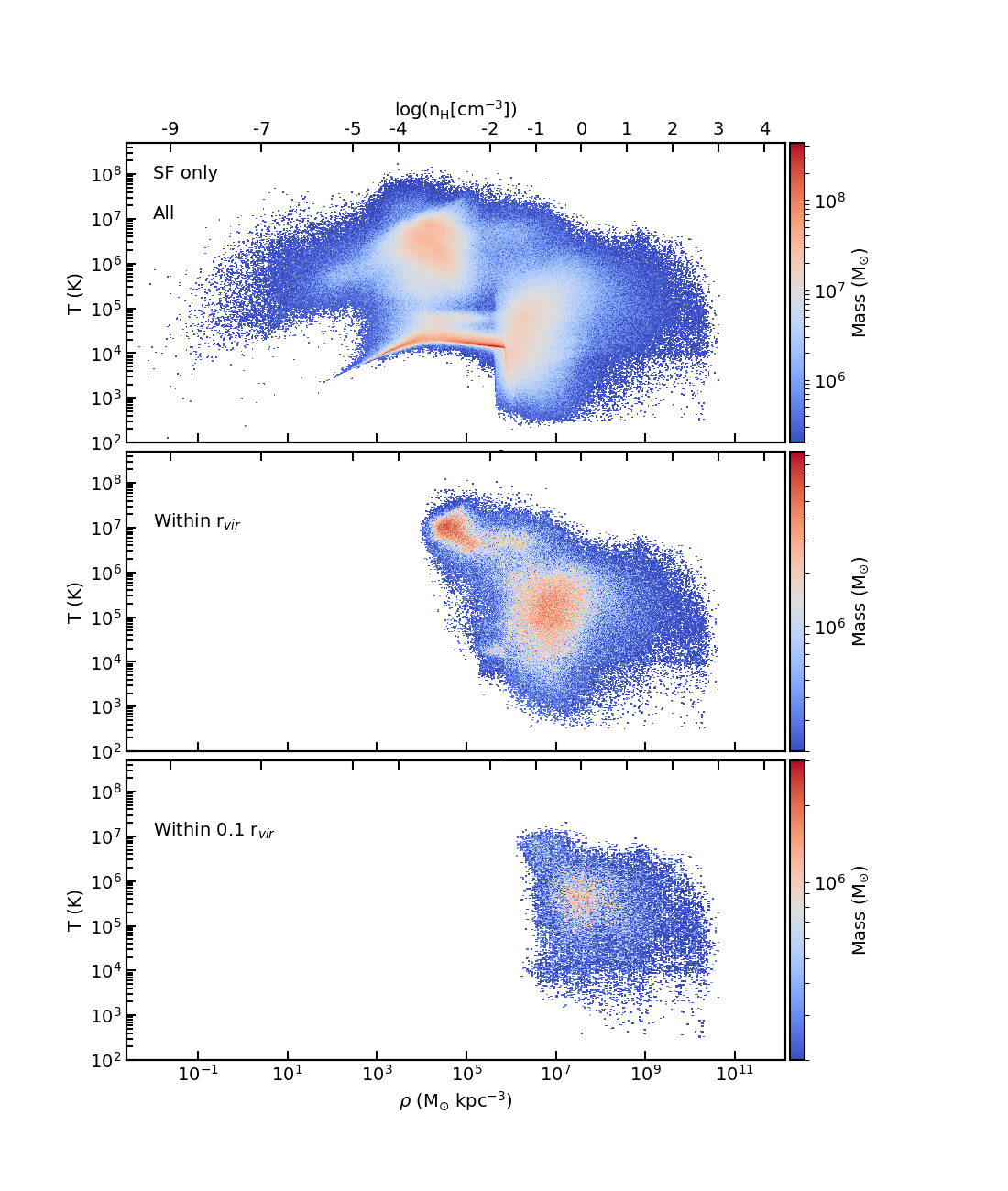} 
\includegraphics[trim=1.8cm 1.5cm 3.5cm 2.5cm, clip, width=.49\textwidth]{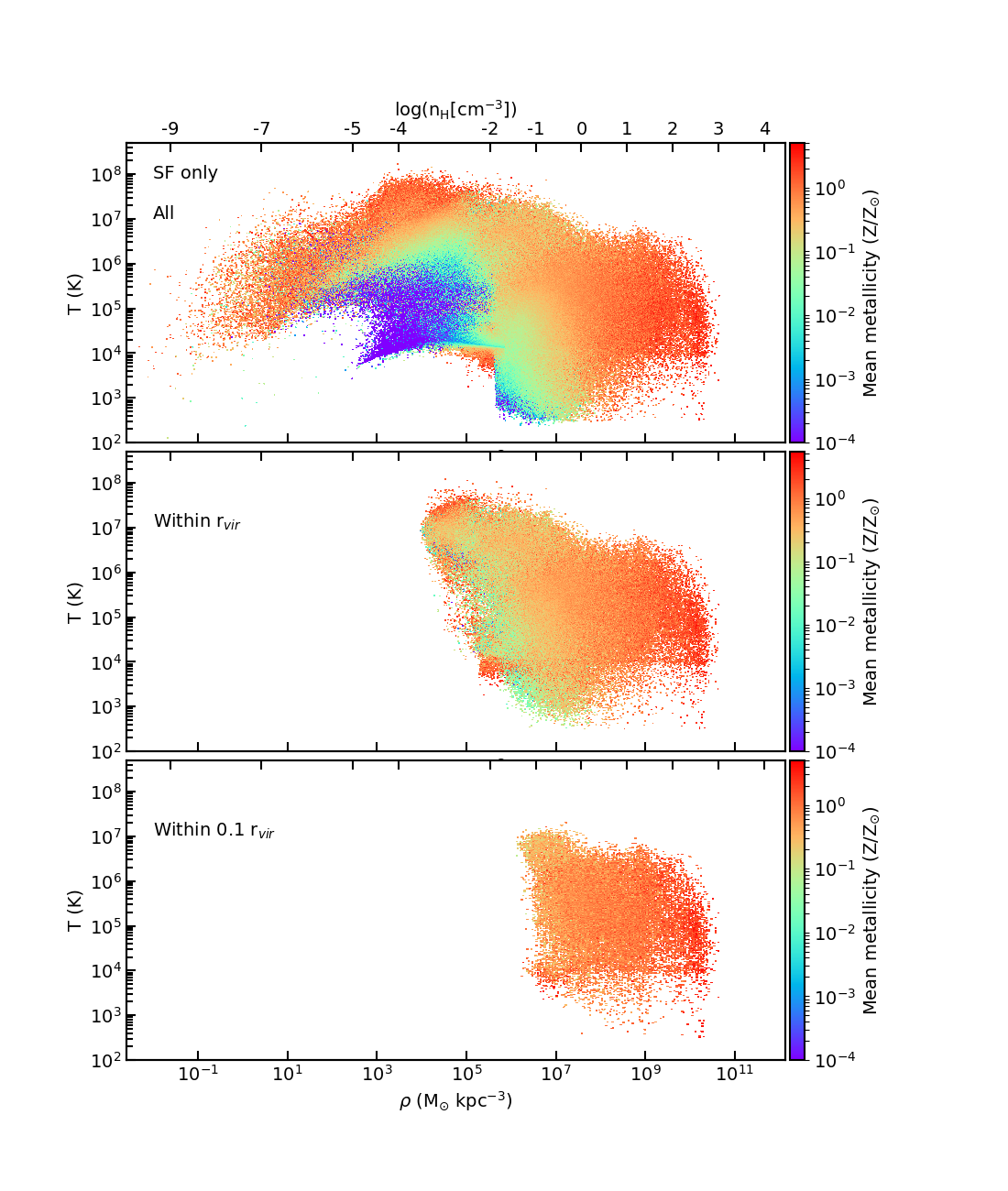}
\end{minipage} 
\caption{Distribution of gas particles in the density-temperature plane 
	in the reference simulation {\sl SF\_only}, at redshift $z = 6$. 
	Top panels show the distribution of all the gas particles in the Lagrangian region,
	middle and bottom panels refer to gas particles within the virial radius r$_{\rm vir}$ and 
	within $0.1$~r$_{\rm vir}$, respectively. 
	The colour encodes the gas mass per density-temperature bin {\sl (left-hand panels)} and 
	the mean metallicity per bin {\sl (right-hand panels)}.
	All the color bars in the left set of panels share the minimum value, while the maximum of the color scale 
	is independent for each panel, to better capture features (the same is true for the three panels on the right). 
	Same as Figure~\ref{PDs}, but for the simulation {\sl SF\_only}.}
\label{PDs_SFonly} 
\end{figure*}

\begin{figure}
\newcommand{\captionfonts}{\small}
\begin{minipage}{\linewidth}
\centering
\includegraphics[trim=0.5cm 0.cm 0.5cm 0.cm, clip, width=1.\textwidth]{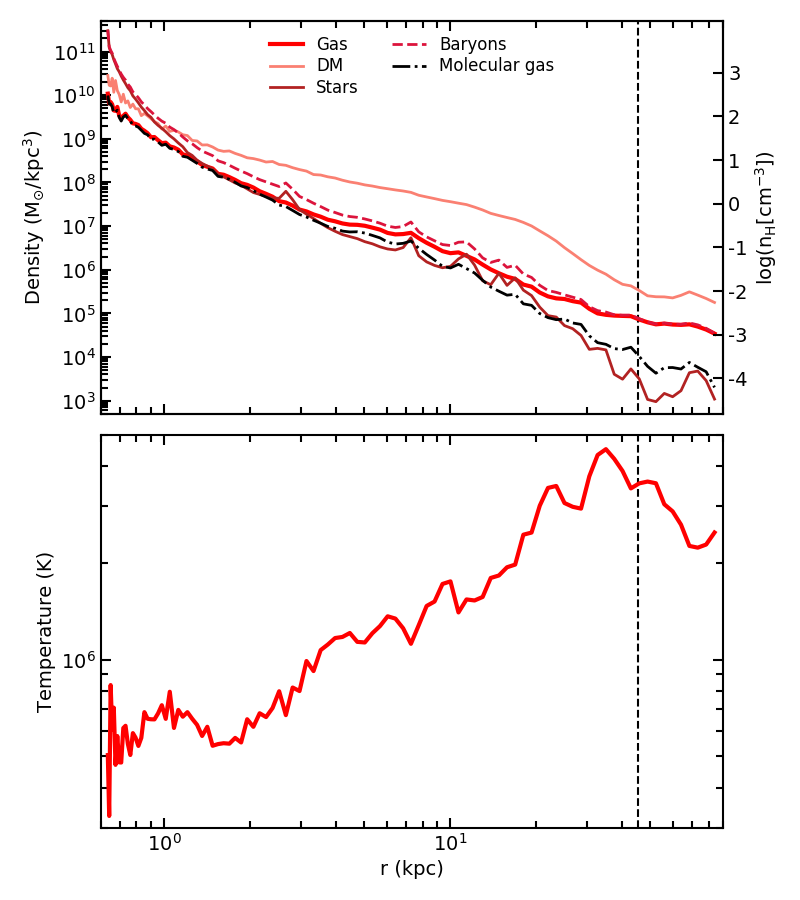} 
\end{minipage} 
\caption{Density and temperature radial profiles  
	within twice the virial radius 
	in the simulation {\sl SF\_only}, at redshift $z = 6$.
	We show the density profile of gas (total and molecular), stars, DM, and baryons {\sl (top panel)}, 
	and the mass-weighted, temperature profile {\sl (bottom panel)}.
	The vertical, dashed black line highligths the virial radius of the most massive subhalo. 
	Same as Figure~\ref{Profiles}, but for the simulation {\sl SF\_only}.}
\label{Profiles_SFonly} 
\end{figure}

In this Appendix we introduce Table~\ref{tab:otherSims}, where we list the most relevant features of central galaxies and 
their most massive BHs for the suite of simulations that we performed to prepare this work. 
It also shows the parameters of the sub-resolution model that we varied (columns~$2 - 6$), 
and how they impact on final properties of central galaxies and their BHs (columns~$7 - 12$), at $z=6$.

In what follows, we focus on the simulation~{\sl SF\_only} and show some further features, 
to ease the comparison with the reference simulation {\sl AGN\_~fid} discussed in Section~\ref{sec:results}. 

Figure~\ref{IntroMapsZoom_SFonly} shows a zoom-in on the central galaxy of the simulation~{\sl SF\_only}, 
while Figure~\ref{PDs_SFonly} illustrates the mass (left panels) and metallicity (right panels) distribution 
in the density-temperature phase diagram of gas particles in the same simulation, at redshift $z = 6$. 
Figure~\ref{Profiles_SFonly} shows density and temperature radial profiles for the aforementioned simulation.


\section*{Data availability}
The data underlying this article will be shared on reasonable request to the corresponding author.


\section*{Acknowledgments}
We thank the anonymous referee for the prompt and careful report 
that helped improving the presentation of results.  
We are grateful to Volker Springel for making the GADGET3 code available to us. 
We thank Klaus Dolag, Stefano Borgani, Giuseppe Murante, and Stefano Carniani for insightful discussions. 
MV is supported by the Excellence Cluster ORIGINS, which is funded by the Deutsche Forschungsgemeinschaft 
(DFG, German Research Foundation) under Germany's Excellence Strategy - EXC-2094 - 390783311. 
MV also acknowledges support from the Alexander von Humboldt Stiftung and 
from the Carl Friedrich von Siemens Stiftung.  
Simulations were carried out using computational resources at Scuola Normale Superiore, 
and Galileo at CINECA (CPU time has been assigned through the project 
"Witnessing the growth of high-redshift SMBHs", PI: M. Valentini). 
We acknowledge the computing centre of Cineca and INAF (under the coordination of the "Accordo Quadro MoU per lo svolgimento di attivit\`{a} congiunta di ricerca - Nuove frontiere in Astrofisica: HPC e Data Exploration di nuova generazione"), for the availability of computing resources and support. 
AF acknowledges support from the ERC Advanced Grant INTERSTELLAR H2020/740120. Any dissemination of results must indicate that it reflects only the author's view and that the Commission is not responsible for any use that may be made of the information it contains. Partial support from the Carl Friedrich von Siemens-Forschungspreis der Alexander von Humboldt-Stiftung Research Award is kindly acknowledged.



\bibliographystyle{mnras} 
\bibliography{cool_ref}


\bsp	
\label{lastpage}
\end{document}